%% file: main.tex
\documentclass{ieeeaccess}
\usepackage[nocompress]{cite}
\usepackage{xcolor}
\usepackage{caption}

\usepackage{spotcolor}
\usepackage{amsmath,amssymb,amsfonts}
\usepackage{algorithmic}
\usepackage{graphicx}
\usepackage{textcomp}
\usepackage{subcaption}
\usepackage{threeparttable}
\usepackage{pict2e}
\usepackage{parskip}
\usepackage{multirow}
\usepackage{colortbl}
\usepackage{algorithmic}
\usepackage{graphicx}
\usepackage{textcomp}
\usepackage{xcolor}
\usepackage{url}
\usepackage{tabularx}
\usepackage{adjustbox}
\usepackage{nameref}
\usepackage{longtable}
\usepackage{threeparttable}
\usepackage{supertabular}
\usepackage{booktabs}
\usepackage{colortbl}
\usepackage{amsmath}
\usepackage{footnote}
\usepackage{siunitx}

\newsavebox{\ORCIDlogo}
\savebox{\ORCIDlogo}{%
\setlength{\unitlength}{\dimexpr 1em/256\relax}%
\begin{picture}(256,256)%
  \color[HTML]{A6CE39}\put(128,128){\circle*{256}}%
  \color{white}%
  \put(78.6,199.2){\circle*{20}}%
  \moveto(70.9,176,9)\lineto(86.3,176,9)\lineto(86.3,69.8)\lineto(70.9,69.8)%
  \closepath\fillpath%
  \moveto(108.9,176.9)\lineto(150.5,176.9)%
  \curveto(190.1,176.9)(207.5,148.6)(207.5 ,123.3)%
  \curveto(207.5,95,8)(186,69.7)(150.7,69.7)%
  \lineto(108.9,69.7)%
  \closepath\fillpath%
  \color[HTML]{A6CE39}%
  \moveto(124.3,83.6)\lineto(148.8,83.6)%
  \curveto(183.7,83.6)(191.7,110.1)(191.7,123.3)%
  \curveto(191.7,144.8)(178,163)(148,163)%
  \lineto(124.3,163)%
  \closepath\fillpath%
\end{picture}%
}
\DeclareSIUnit\TWh{TWh}
\DeclareSIUnit[number-unit-product = {}]{\percent}{\%}
\definecolor{myblue}{HTML}{004392}
\newcommand\orcidicon[1]{\raisebox{0.4ex}{\usebox{\ORCIDlogo}}}
\usepackage[hidelinks,colorlinks=true, linkcolor=blue, citecolor=blue, urlcolor=black]{hyperref}

\def\BibTeX{{\rm B\kern-.05em{\sc i\kern-.025em b}\kern-.08em
    T\kern-.1667em\lower.7ex\hbox{E}\kern-.125emX}}
\begin{document}
\bstctlcite{IEEEtranBSTCTL}
\history{Received 8 October 2025, accepted 19 November 2025.}
\doi{10.1109/ACCESS.2025.3636266}
\title{Efficient Kernel Mapping and Comprehensive System Evaluation of LLM Acceleration on a CGLA}

\author{\uppercase{Takuto ANDO}\href{https://orcid.org/0009-0005-1873-869X}{\orcidicon{0009-0005-1873-869X}}, \IEEEmembership{Member, IEEE},
\uppercase{Yu ETO}\href{https://orcid.org/0009-0006-0640-1683}{\orcidicon{0009-0006-0640-1683}},\\ \uppercase{Ayumu TAKEUCHI}\href{https://orcid.org/0009-0006-2094-3163}{\orcidicon{0009-0006-2094-3163}} and \uppercase{Yasuhiko NAKASHIMA}\href{https://orcid.org/0000-0002-9457-5061}{\orcidicon{0000-0002-9457-5061}}, \IEEEmembership{Member, IEEE}}

\address{Nara Institute of Science and Technology (NAIST), Ikoma, Nara 630-0192, Japan}

\tfootnote{This work was supported by the JST-ALCA-Next Program (Grant Number JPMJAN23F4) and JSPS KAKENHI (Grant No. 22H00515). We also acknowledge the activities of VDEC, The University of Tokyo, in collaboration with NIHON SYNOPSYS G.K.}

\markboth
{T. Ando \headeretal: Efficient Kernel Mapping and Comprehensive System Evaluation of LLM Acceleration on a CGLA}
{T. Ando \headeretal: Efficient Kernel Mapping and Comprehensive System Evaluation of LLM Acceleration on a CGLA}

\corresp{Corresponding author: Takuto ANDO (e-mail: antaku7585@gmail.com).}

\input{./abstract.tex}

\titlepgskip=-15pt

\maketitle

\input{./introduction.tex}

\input{./related_work.tex}

\input{./proposed.tex}

\input{./experiments_and_results.tex}

\input{./discussion.tex}

\input{./conclusion.tex}

\section*{Data Availability}
To ensure the reproducibility of our results and to support further research in this area, the source code, kernel implementations, build scripts, and experimental manifests used in this work will be made publicly available upon publication of this article.
The artifacts will be accessible at the following repository: \url{https://github.com/Takuto-Ando/IMAX3-LLM}.

\bibliographystyle{IEEEtran}
\bibliography{bibliography}
\vspace{50em}
\begin{IEEEbiography}[{\includegraphics[width=1in,height=1.25in,clip,keepaspectratio]{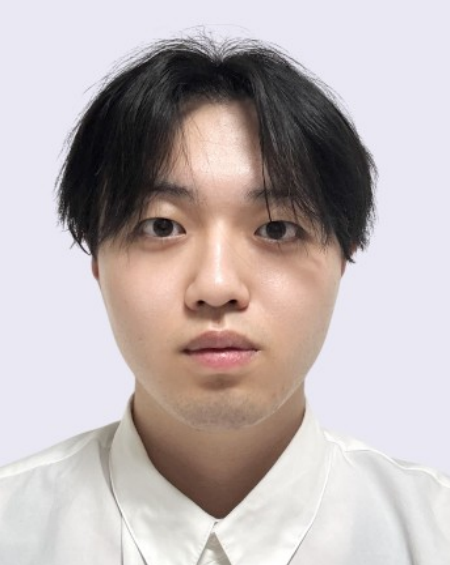}}]{TAKUTO ANDO}(Member, IEEE) received the B.E. degree in engineering from the National Institute of Technology, Oita College, Japan, in 2025. 
He is currently pursuing the M.E. degree with the Graduate School of Science and Technology, Nara Institute of Science and Technology (NAIST). 
His research interests include computer architecture, circuit design, domain-specific architecture, and machine learning applications.
He is a Student Member of IPSJ and IEICE.
\end{IEEEbiography}
\vspace{-3em}
\begin{IEEEbiography}[{\includegraphics[width=1in,height=1.25in,clip,keepaspectratio]{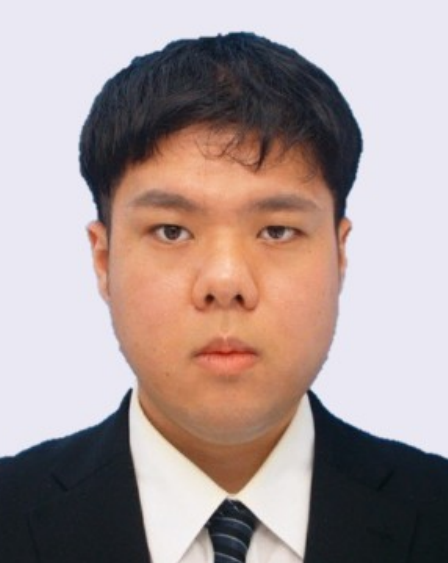}}]{YU ETO} received the B.E. degree in engineering from the National Institute of Technology, Nara College, Japan, in 2024. 
  He is currently pursuing the M.E. degree with the Graduate School of Science and Technology, Nara Institute of Science and Technology (NAIST). 
  His research interests include computer architecture, emulation, circuit design, design for testability, asynchronous circuits and accelerators.
\end{IEEEbiography}
\vspace{-3em}
\begin{IEEEbiography}[{\includegraphics[width=1in,height=1.25in,clip,keepaspectratio]{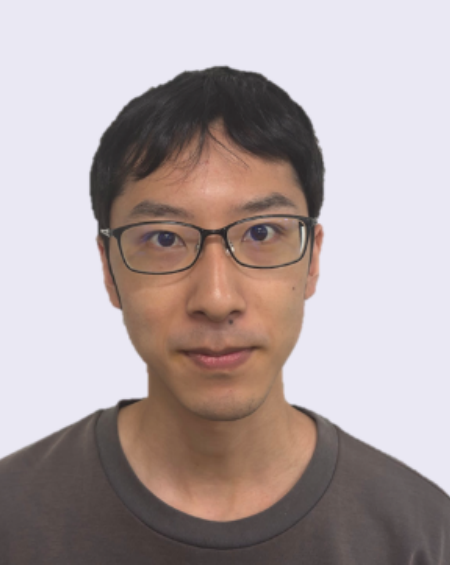}}]{Ayumu takeuchi} received the B.E. degree in engineering from the National Institute of Technology, Kagawa College, Japan, in 2024. 
  He is currently pursuing the M.E. degree with the Graduate School of Science and Technology, Nara Institute of Science and Technology (NAIST). 
  His research interests include computer architecture, attention mechanisms, and large language models. 
  \end{IEEEbiography}
\vspace{-3em}
\begin{IEEEbiography}[{\includegraphics[width=1in,height=1.25in,clip,keepaspectratio]{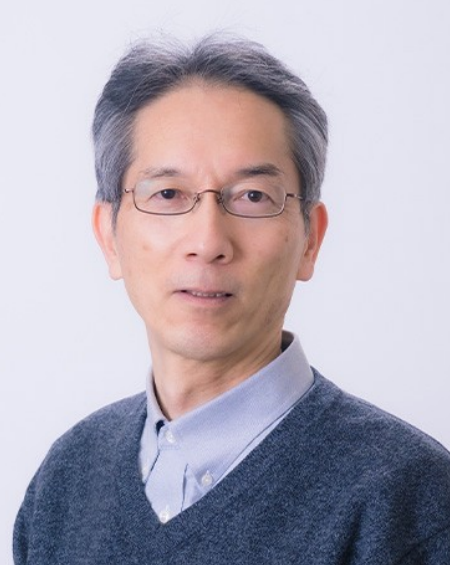}}]{YASUHIKO NAKASHIMA} 
  (Senior Member, IEEE) received the B.E., M.E., and Ph.D. degrees in computer engineering from Kyoto University, in 1986, 1988, and 1998, respectively. 
  He was a Computer Architect with the Computer and System Architecture Department, FUJITSU Ltd., from 1988 to 1999. From 1999 to 2005, he was an Associate Professor with the Graduate School of Economics, Kyoto University. 
  Since 2006, he has been a Professor with the Graduate School of Information Science, Nara Institute of Science and Technology (NAIST). 
  His research interests include computer architecture, emulation, circuit design, and accelerators. He is a fellow of IEICE.
\end{IEEEbiography}

\EOD

\end{document}

%% file: abstract.tex
\begin{abstract}

    Large Language Models (LLMs) demand substantial computational resources, resulting in high energy consumption on GPUs.
    To address this challenge, we focus on Coarse-Grained Reconfigurable Arrays~(CGRAs) as an effective alternative that provides a trade-off between energy efficiency and programmability.
    This paper presents the first comprehensive, end-to-end evaluation of a non-AI-specialized Coarse-Grained Linear Array~(CGLA) accelerator for the state-of-the-art Qwen3 LLM family. 
    The architecture has a general-purpose, task-agnostic design, yet its flexible instruction set allows for domain-specific adaptations.
    This flexibility enables the architecture to achieve high efficiency for sustainable LLM inference.
    We assess the performance of our architecture on an FPGA prototype using the widely adopted llama.cpp framework. 
    We then project its potential as a \SI{28}{\nano\meter} ASIC and compare it against a high-performance GPU (NVIDIA RTX 4090) and an edge AI device (NVIDIA Jetson AGX Orin).
    While GPUs exhibit lower latency, our non-AI-specific accelerator achieves higher energy efficiency, improving the Power-Delay Product (PDP) by up to \num{44.4}$\times$ and \num{13.6}$\times$ compared with the RTX 4090 and Jetson, respectively.
    Similarly, it reduces the Energy-Delay Product~(EDP) by up to \num{11.5}$\times$ compared to the high-performance GPU, demonstrating a favorable performance-energy trade-off.
    Critically, our system-level analysis identifies host-accelerator data transfer as the primary performance bottleneck, a factor often overlooked in kernel-level studies.
    These findings provide design guidance for next-generation LLM accelerators.
    This work validates CGRAs as a suitable platform for LLM inference in power-constrained environments, without being confined to specific algorithms.
    
\end{abstract}

\begin{keywords}
Qwen, LLM, CGRA, CGLA, IMAX
\end{keywords}

%% file: introduction.tex
\section{Introduction}
Large Language Models~(LLMs) have evolved from specialized text processing tools into a widely applied technology in various fields\nobreak\cite{LLM_survey1,LLM_survey2,LLM_survey3,LLM_survey4,LLM_survey5}.
They have been applied in fields such as natural language processing\nobreak\cite{NLP_survey1,NLP_survey2,NLP_survey3}, code generation\nobreak\cite{code_generation1,code_generation2,code_generation3}, and conversational AI\nobreak\cite{conversational_agents1,conversational_agents2}.
The development of multimodal LLMs, which integrate modalities such as text, images, and audio, is further expanding their scope of application on a large scale\nobreak\cite{multi-modal1,multi-modal4,multi-modal5}. 
The interaction between algorithmic development and computational architectures enables these innovations.
Therefore, improving this infrastructure is not merely a practical challenge but critical for the broader societal integration of AI.
Currently, this technological progress primarily depends on the evolution of General-Purpose Graphics Processing Units~(GPGPUs).
However, despite these advances, achieving high performance with LLMs requires substantial computational resources for both training and inference.
In particular, the high performance of GPGPUs requires substantial power, which creates significant environmental and economic concerns\nobreak\cite{gpu_power1,gpu_power2}. 
The International Energy Agency~(IEA) estimates that data center electricity consumption could double by 2030, reaching approximately~\SI{945}{\TWh}\nobreak\cite{iea_energy_ai}. 
This consumption level, fueled primarily by the expanding demand for AI, slightly exceeds Japan's total annual electricity consumption.
This sustainability challenge presents a significant barrier to the widespread adoption of LLM technology, making it necessary to address this with improvements in computational architecture.

To address this challenge, specialized hardware such as Application-Specific Integrated Circuits~(ASICs) and Field-Programmable Gate Arrays~(FPGAs) have been widely investigated.
While edge GPUs, such as the NVIDIA Jetson series, demonstrate progress in reducing power consumption, their general-purpose graphics pipelines inherently constrain substantial gains in power efficiency.
In contrast, ASICs can potentially achieve orders of magnitude higher power efficiency. 
ASICs specialize their circuits for the core computational patterns of LLM inference, such as dot-product operations. 
This allows them to remove all unnecessary functionality and achieve maximum performance for a specific task.
However, this high efficiency is intrinsically coupled with a lack of flexibility, rendering these designs unable to adapt to evolving algorithms.

Therefore, we consider Coarse-Grained Reconfigurable Arrays~(CGRAs) as an architectural paradigm that addresses the fundamental trade-off between efficiency and flexibility.
CGRAs achieve power efficiency approaching that of an ASIC by mapping dataflow graphs directly onto an array of processing elements~(PEs), while preserving programmability through their reconfigurable fabric.
This work focuses on IMAX\nobreak\cite{imax3}, a general-purpose accelerator based on Coarse-Grained Linear Arrays~(CGLAs), a type of CGRA architecture.
IMAX features a linear arrangement of PEs and Local Memory Modules~(LMMs), a design specifically intended to efficiently handle the irregular memory access patterns of LLM inference.

Our previous work has demonstrated the versatility and effectiveness of the IMAX architecture across a diverse range of workloads, including Sparse General Matrix Multiplication~(SpGEMM), Fast Fourier Transform~(FFT), and Convolutional Neural Networks~(CNNs)\nobreak\cite{imax3,cnn_imax_1,cnn_imax_2}. 
We have also extended this versatility to LLMs, showing the feasibility of implementing key kernels from the Llama2-based model on IMAX\nobreak\cite{first_llm_imax,llama2_imax}. 
Based on this foundation, this work provides the first comprehensive evaluation of IMAX as an LLM accelerator.
Specifically, we expand our analysis to the state-of-the-art Qwen3 model\nobreak\cite{yang2025qwen3technicalreport,qwen}. 
We adopt the widely used C/C++ inference engine llama.cpp\nobreak\cite{llama.cpp} to assess performance in practical scenarios.
This approach allows our evaluation to encompass the entire system from end-to-end (E2E), rather than being limited to single-kernel performance.
This evaluation accounts for host-CPU interaction, data transfer overheads, and complex control flows.

The main contributions of this paper are as follows:
\begin{itemize}
    \item We present the first comprehensive, E2E evaluation of a general-purpose CGRA for modern LLM inference. Our work demonstrates that a non-AI-specialized architecture can achieve superior energy efficiency, improving the Power-Delay Product (PDP) by up to \num{44.4}$\times$ and the Energy-Delay Product (EDP) by up to \num{11.5}$\times$ compared to a high-performance GPU. This finding validates a sustainable hardware approach that avoids the risk of rapid obsolescence associated with task-specific ASICs.
    \item We identify a critical performance bottleneck shift from computation to host-accelerator data transfer. Our system-level analysis using the llama.cpp framework reveals that the decode phase is data-transfer-bound, a finding obscured in kernel-centric evaluations that provides crucial guidance for latency optimization.
    \item We validate CGRAs as a viable architectural solution that balances efficiency and programmability for sustainable LLM inference. Our work establishes that general-purpose reconfigurable hardware can effectively adapt to rapid algorithmic evolution, making it a practical choice for power-constrained deployment.
\end{itemize}

The remainder of this paper is organized as follows. 
Section~\ref{rwork} surveys related work in LLM acceleration and introduces the IMAX architecture.
Section~\ref{proposed} details our implementation methodology, including the execution framework and kernel mapping strategies.
Section~\ref{ex_and_re} presents the experimental results, comparing our accelerator against GPUs on E2E latency and energy-efficiency metrics.
Section~\ref{discussion} analyzes performance bottlenecks by examining on-chip memory impact, execution phase breakdowns, and host system limitations, and discusses directions for future work.
Finally, Section~\ref{conclusions} concludes the paper.

%% file: related_work.tex
\section{Related Work}
\label{rwork}
This work relates to both hardware acceleration for LLMs and CGRA implementation.
In this section, we survey the key trends in both domains to contextualize our work and clarify its unique position and contributions.

\subsection{LLM Hardware Accelerators}
Research on hardware acceleration for LLM inference falls into three main streams. 
The primary stream centers on the use of GPGPUs.
Innovations such as Tensor Cores and the Hopper architecture\nobreak\cite{nvidia_h100} have substantially increased the throughput of LLM inference.
The high programmability and well-developed ecosystem of GPGPUs facilitate rapid prototyping and application development\cite{sze2017efficientprocessingdeepneural}.
However, this approach consumes substantial power, posing a significant sustainability challenge\cite{power_consideration}.

The second stream involves the development of ASICs specifically tailored for LLM inference. 
Architectures such as Google's TPU\nobreak\cite{TPU} and Meta's MTIA\nobreak\cite{MTIA} achieve performance and power efficiency that far surpass GPGPUs by concentrating computational resources on the large-scale dot-product operations at the core of LLMs. 
However, this high efficiency often results in over-specialization to particular algorithms or quantization methods, limiting adaptability.
Consequently, these designs inherently risk rapid obsolescence due to the difficulty in adapting to fast-evolving LLM algorithms such as Mixture of Experts~(MoE)\nobreak\cite{MoE} and Activation-aware Weight Quantization~(AWQ)\nobreak\cite{AWQ}.

The third stream, positioned between the efficiency of ASICs and the flexibility of GPGPUs, explores the use of reconfigurable hardware, such as FPGAs and CGRAs. 
These architectures have already proven effective for general AI acceleration, with numerous research demonstrating their advantages in balancing performance and power consumption~\cite{fpgacnn,li2020ftransenergyefficientaccelerationtransformers,vitisai,multi_task}. 
Based on this success, they are now emerging as a suitable solution for LLM inference. 
The reconfigurability of these devices can potentially address the inflexibility inherent in ASICs, offering a suitable approach for the rapid evolution of LLMs. 

\subsection{FPGA for LLM Inference}

Research on FPGA-based LLM acceleration is an important focus of current research in algorithm-hardware co-design\nobreak\cite{hydra,llm_agile,llm4fpga,llama2fpga}. 
For instance, UltraFormer\nobreak\cite{ultraformer} exemplifies this approach by redesigning the Transformer architecture itself for FPGAs. 
It pursues algorithmic-level optimization by integrating a linear-complexity attention mechanism with \SI[number-unit-product=\nobreakdash-]{1.58}{bit} quantization derived from BitNet\nobreak\cite{bitnet}. 
Similarly, research like BitMoD\nobreak\cite{bitmod} and AccLLM\nobreak\cite{acc_llm} have demonstrated a method for improving performance and efficiency while maintaining model accuracy. 
They achieve this by proposing custom low-bit data types, such as \SI[number-unit-product=\nobreakdash-]{2}{bit} and \SI[number-unit-product=\nobreakdash-]{3}{bit} formats, and implementing specialized arithmetic units on FPGAs to process them efficiently.

Furthermore, FPGAs provide a suitable platform for experimenting with innovative architectures to address the memory bandwidth bottleneck.
For example, FlexCiM\nobreak\cite{flex_cim} leverages fully digital compute-in-memory (DCiM) technology to support flexible structured sparsity, while MECLA\nobreak\cite{mecla} substantially reduces off-chip data transfers through a novel on-chip weight reorganization technique. 
The research area is also expanding beyond single-accelerator designs to establish development ecosystems.
A notable example is SECDA-LLM\nobreak\cite{secda-llm}, which proposes a framework for rapidly integrating FPGA accelerators within llama.cpp. 
Although FPGAs are an active area of research for LLM acceleration, they still encounter challenges such as high development complexity and inherent limitations in operating frequency and power efficiency compared to ASICs.

\begin{figure*}[t]
    \centering
    \includegraphics[width=1.95\columnwidth]{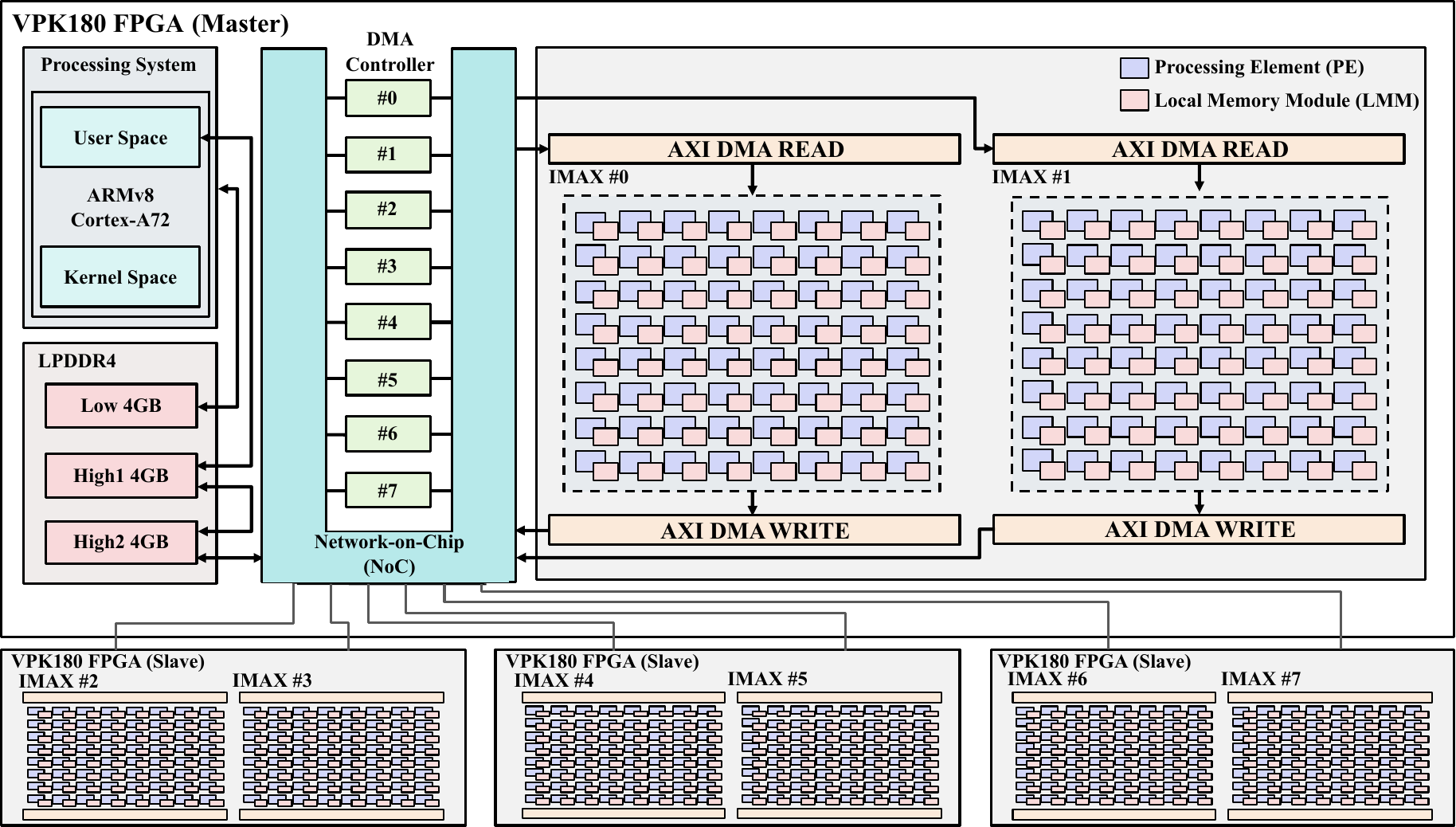}
    \caption{ High-level overview of the IMAX3 system architecture, implemented on a multi-FPGA platform with four AMD Versal VPK180 devices.}
    \label{fig:imax3_conf}
  \end{figure*}

\subsection{CGRA for LLM Inference}
CGRAs have recently been investigated as an architectural solution to overcome the performance and programmability trade-offs of FPGAs. 
By spatially mapping dataflow graphs~(DFGs) onto an array of PEs, CGRAs achieve power efficiency approaching that of ASICs while retaining programmability through reconfiguration. 
Research on CGRAs for LLM acceleration is an emerging field, with several distinct approaches recently introduced.

Other research focuses on optimizing linear operations like Matrix-Vector Multiplication~(MVM), which represent a major computational component in LLMs. 
The CORO algorithm\nobreak\cite{coro}, for instance, exemplifies a data-centric optimization strategy, combining non-uniform quantization with variable-length data encoding. 
This technique leverages the clustering of weight values to reduce the memory footprint by encoding them in pairs. 
CORO's contribution lies in improving algorithmic memory efficiency for kernel-level performance on flexible accelerators. 
However, it does not extend to broader architectural and system-level challenges.

Another research direction focuses on the optimization of non-linear operations. 
While these operations account for a smaller portion of the total computation compared to MVM, they often become significant latency bottlenecks. 
A representative work in this area is PICACHU\nobreak\cite{picachu}, a plug-in CGRA accelerator designed to efficiently handle diverse non-linear functions such as Softmax and normalization. 
It is designed to complement existing LLM accelerators by offloading only non-linear computations to the CGRA.

Our work differs from these works in two key aspects: its architectural philosophy and its evaluation methodology. 
First, regarding our architectural philosophy, we emphasize architectural versatility. 
Similar to CORO, our work focuses on offloading dot-product operations, which account for the majority of the computational load in LLMs. 
In contrast to CORO, which is designed specifically for dot-product operations in LLMs, our IMAX architecture is a general-purpose CGRA that is not limited to a single task or model.
We have already demonstrated its capability on diverse workloads such as FFT\nobreak\cite{imax3} and CNNs\nobreak\cite{cnn_imax_1,cnn_imax_2}.
This research shows that the same architecture can achieve high energy efficiency on modern LLM inference tasks without modification, demonstrating a design principle for sustainable hardware that avoids task-specific specialization.
A second key difference is our evaluation methodology. 
We assess E2E system performance using the widely used llama.cpp framework. 
This approach enables the assessment of E2E system performance under realistic conditions. 
Our analysis incorporates important factors such as host-CPU interaction, DMA transfer overheads, and complex software control. 
In the emerging field of LLM acceleration on CGRAs, this paper provides a comprehensive characterization of the performance and challenges of a versatile architecture at a full-system level.

While a direct quantitative comparison with other FPGA and ASIC-based LLM accelerators is valuable, achieving a fair comparison is challenging.
This is due to discrepancies in target models, quantization schemes, process technologies, and evaluation methodologies.
For instance, some studies use isolated kernel benchmarks, whereas others perform E2E system-level measurements.
Many papers report performance in TOPS or TOPS/W based on peak theoretical throughput, figures that often omit significant system-level overheads.
Therefore, this work provides a detailed performance analysis of a general-purpose, non-specialized CGRA architecture, using the industry-standard llama.cpp. 
The C++ framework is used unmodified, and we compare our results against those of GPU platforms. 
Our evaluation inherently incorporates host-accelerator interactions and data transfer overheads, thereby offering a realistic assessment of system-level performance and energy efficiency.


\subsection{CGLA architecture and IMAX}
The IMAX architecture is based on a different design approach from those previously mentioned.
The architecture is designed for task-agnostic versatility, providing both linear scalability and high programmability through direct compilation from C/C++.
IMAX employs a CGLA structure, arranges PEs and LMMs in an alternating pattern in a one-dimensional array. 
This design addresses the complex routing and compilation challenges of conventional 2D mesh-based CGRAs.
The combination of this simple structure with multi-functional, CISC-based PEs allow IMAX to efficiently execute diverse workloads. 
These include not only linear operations like General Matrix Multiplication~(GEMM) but also those involving irregular memory accesses and complex control flows.
This linear topology is particularly well-suited for the dot-product operations that dominate LLM inference, providing a theoretical basis for its high energy efficiency. 
Specifically, weights and activations can be streamed through the one-dimensional array of PEs, creating a deep, deterministic pipeline. 
This structure improves spatial data locality and reuse. 
Data passed from one PE is immediately consumed by its neighbor, which eliminates the need for complex routing or access to a higher-level memory hierarchy.
The highly regular and predictable memory access patterns inherent in this dataflow execution minimize the energy overhead associated with data movement and address calculation. 
These factors contribute significantly to reducing power consumption in conventional architectures.

\begin{figure}[t]
    \centering
    \includegraphics[width=0.95\columnwidth]{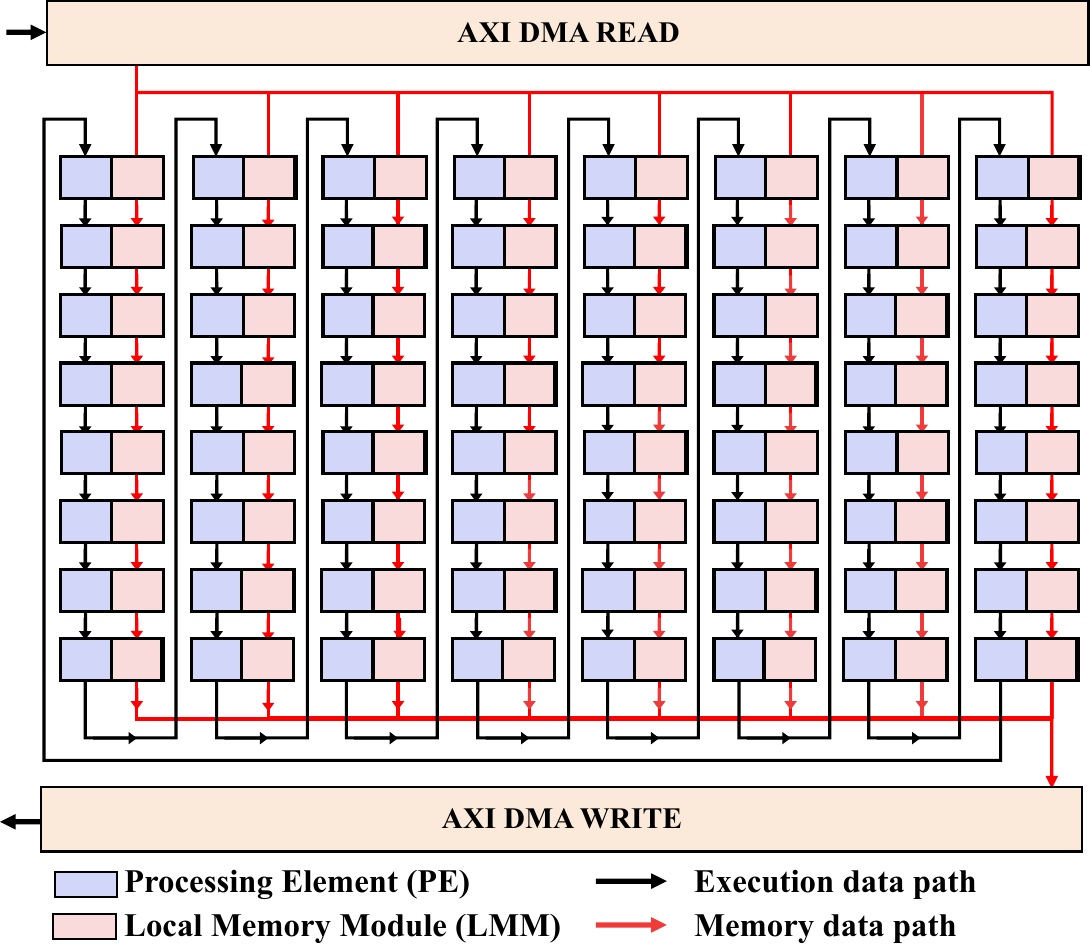}
    \caption{ Inter-PE and LMM connections within a single IMAX compute lane.}
    \label{fig:pe_detailed}
\end{figure}

Fig.~\ref{fig:imax3_conf} provides a high-level overview of the IMAX3 system architecture. 
Implemented on an AMD Versal VPK180 FPGA, IMAX3 is a System-on-Chip~(SoC) consisting of a Processing System~(PS) and Programmable Logic~(PL). 
The PS serves as the host, featuring a dual-core Arm Cortex-A72 processor running Linux. 
It manages the entire workflow, from application compilation to execution control. 
A high-bandwidth Network-on-Chip~(NoC) facilitates communication between the PS and PL, while a dedicated DMA controller efficiently transfers data between system memory and the accelerator.
The accelerator core itself resides within the PL. 
It consists of eight independent compute lanes. 
This multi-lane design is central to the architecture's scalability, as performance can be tailored by allocating a variable number of lanes to match an application's degree of parallelism.

The internal structure of each compute lane is depicted in Fig.~\ref{fig:pe_detailed}. 
This illustrates the core CGLA structure of IMAX, where PEs and LMMs are arranged alternately in a one-dimensional array. 
This linear topology simplifies the data paths between PEs, facilitating easier compilation while also enabling low-latency access to adjacent LMMs. 
Data is loaded into the LMMs via DMA and then forms a pipelined dataflow across the PEs, which is essential for maintaining high utilization of the arithmetic units.

\begin{figure}[t]
    \centering
      \includegraphics[width=0.95\columnwidth]{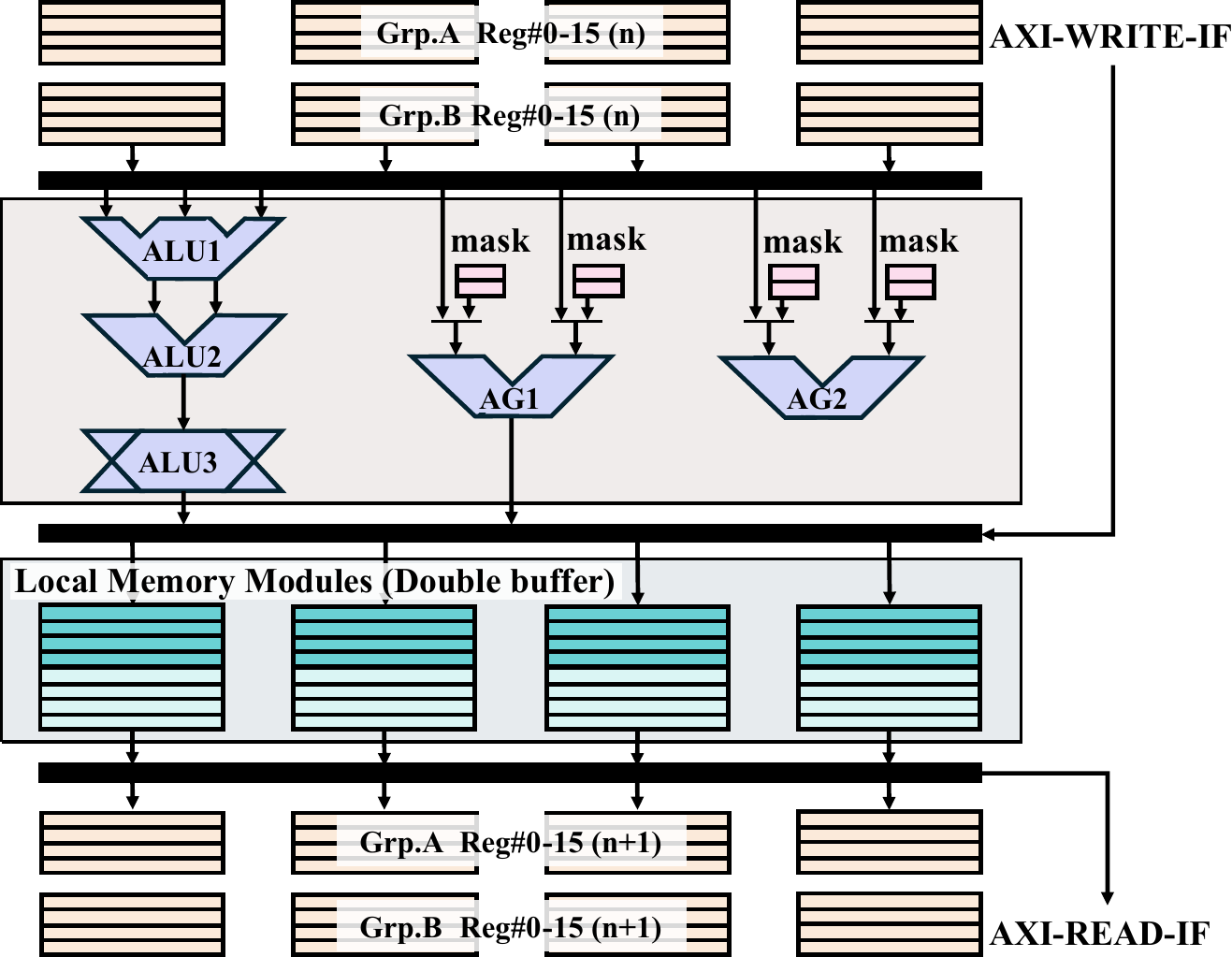}
    \caption{Detailed architecture of a single IMAX PE.}
    \label{fig:execution_unit}
  \end{figure}

Fig.~\ref{fig:execution_unit} details the internal architecture of an individual PE. 
Each PE features a heterogeneous design, comprising multiple arithmetic units, address generation units, and LMMs. 
The arithmetic component consists of three ALUs (ALU1, ALU2, and ALU3), dedicated to integer, logical, and shift operations, respectively, which enables the parallel execution of diverse instructions. 
The Address Generation Units (AG1, AG2) operate independently of the ALUs to calculate memory addresses. 
This design decouples the computation and memory access pipelines, thereby improving execution efficiency. 
The LMM is implemented with a hardware-managed double-buffered configuration, a key feature designed to enable the concurrent execution of computation and data transfers. 
While one buffer is being used by the PEs for active computation, the DMA controller can simultaneously load the next set of data into the other buffer. 
This mechanism is intended to mask memory access latency by overlapping communication with computation, thereby maximizing data throughput.
The integration of these components allows each PE to efficiently process the large volumes of data demanded by high-performance computing tasks.

Previous work has established IMAX3 as a general-purpose computing platform. 
Its effectiveness has been demonstrated across a wide spectrum of workloads. 
These range from traditional kernels, such as SpGEMM, FFT\nobreak\cite{imax3}, light-field image processing\nobreak\cite{light_field_imax} to modern AI applications such as CNNs\nobreak\cite{cnn_imax_1,cnn_imax_2}, Graph Convolutional Networks~(GCNs)\nobreak\cite{gat_imax}, and Retrieval-Augmented Generation (RAG) systems\nobreak\cite{knn_imax}.
This work applies these findings to the domain of LLMs. 
Our previous works\nobreak\cite{first_llm_imax,llama2_imax} presented the implementation of a Llama2-based model on IMAX, confirming its feasibility.
Based on that success, this work aims to validate the performance and robustness of IMAX as an LLM accelerator using a wider range of models and workloads.
To validate the generalizability of the architecture, we evaluate its robustness against the Qwen3 family, which features diverse model structures.
Therefore, we expand our evaluation to the state-of-the-art Qwen3 family to assess its practical performance.

%% file: proposed.tex
\section{LLM Implementation on a CGLA}
\label{proposed}

\begin{figure}[t]
  \centering
  \includegraphics[width=0.95\columnwidth]{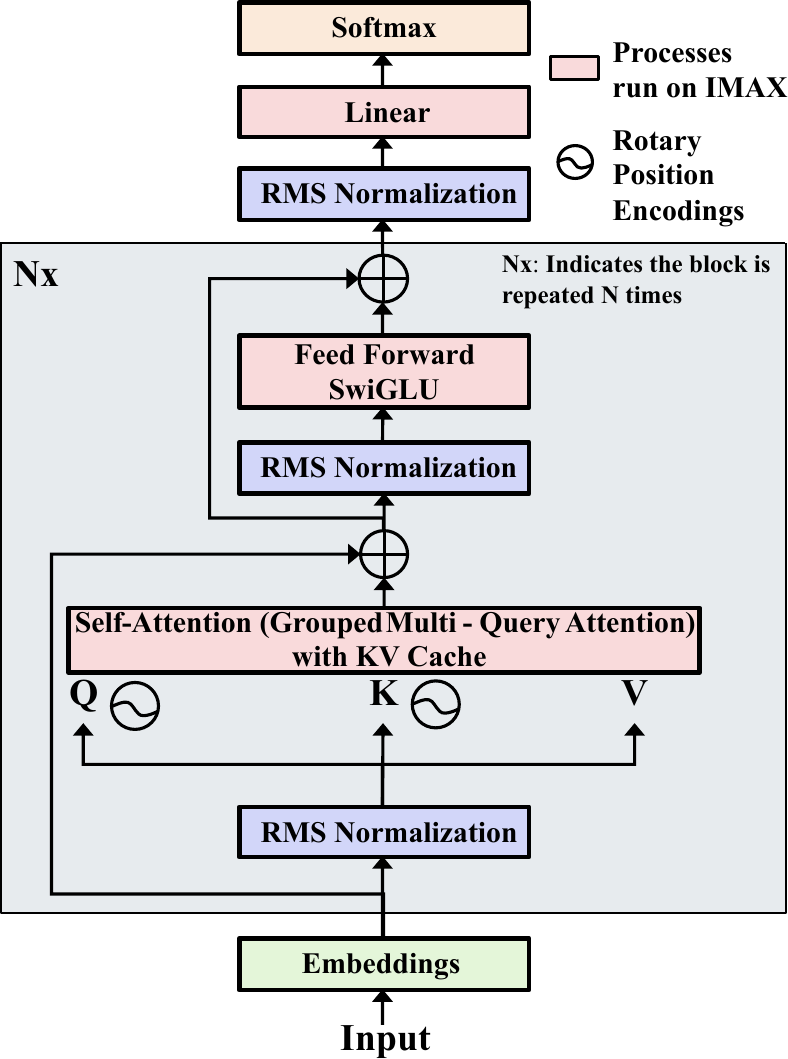}
  \caption{Overview of the LLM inference architecture, based on the llama.cpp framework, and our proposed task partitioning.}
  \label{fig:llama_arch}
\end{figure}

\begin{figure*}[t]
  \centering
  \includegraphics[width=1.85\columnwidth]{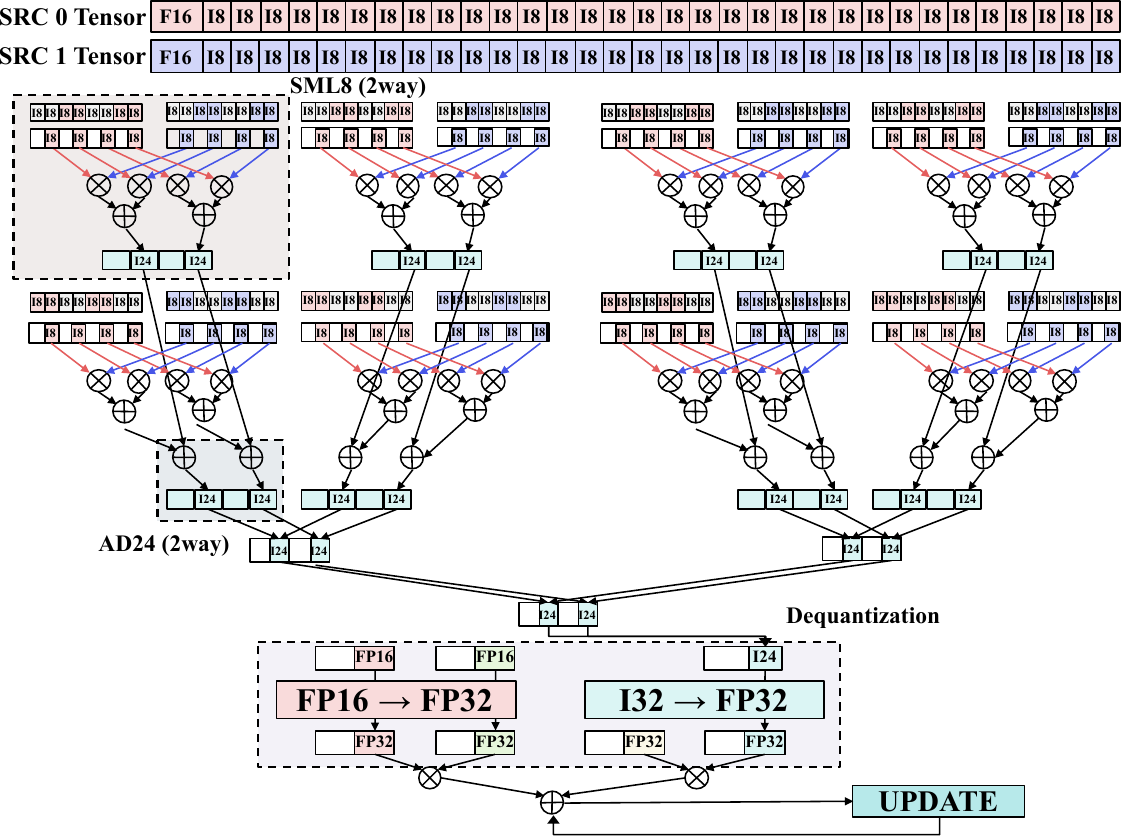}
  \caption{Detailed dataflow diagram for the parallelized Q8\_0 dot-product operation. This kernel uses parallel SML8 and AD24 custom instructions to perform the dot-product operation.}
  \label{fig:q8dot}
  \end{figure*}

  \begin{figure}[t]
  \centering
  \includegraphics[width=0.95\columnwidth]{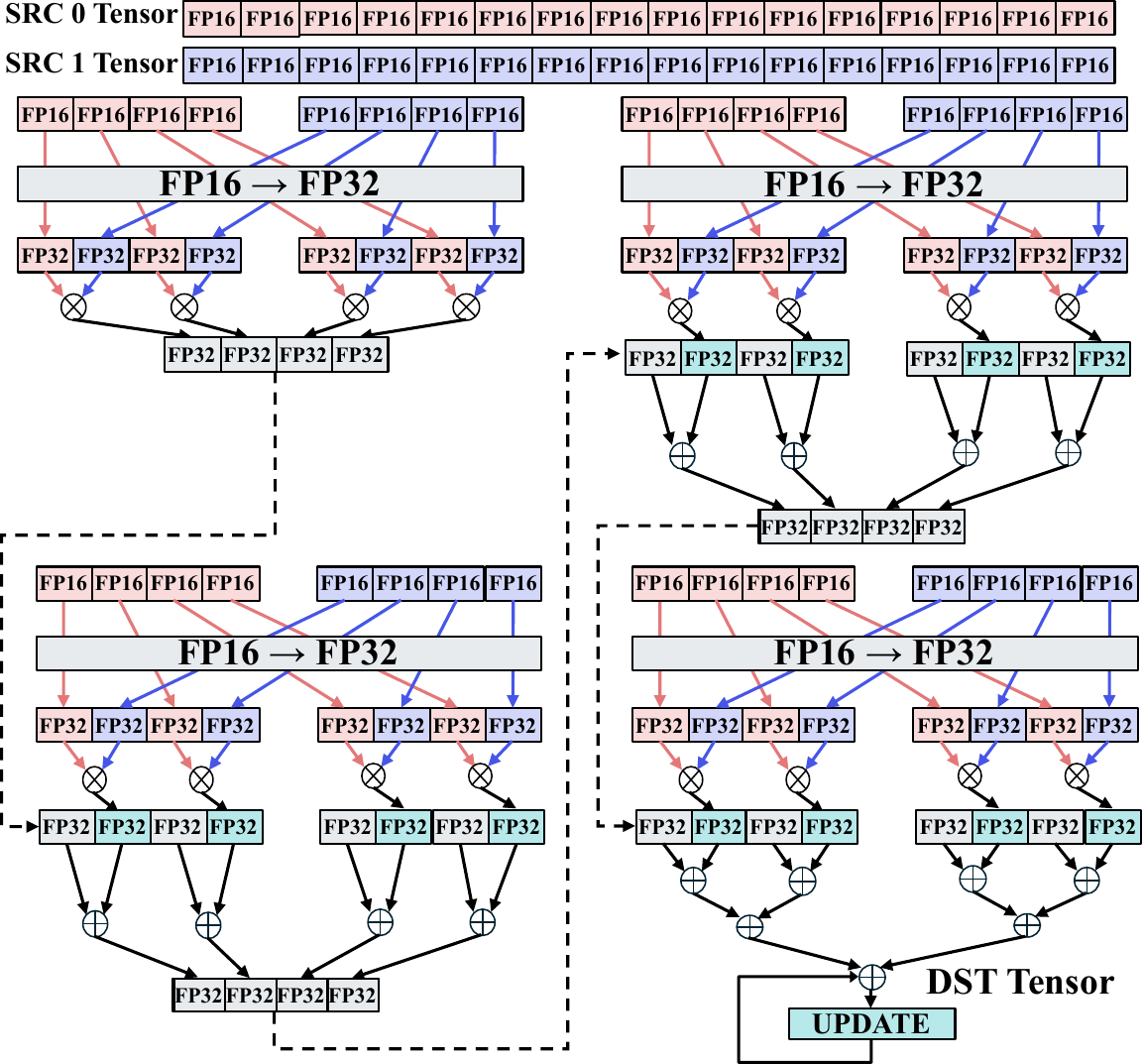}
  \caption{Detailed dataflow diagram for the parallelized FP16 dot-product operation as mapped onto the IMAX architecture.}
  \label{fig:f16dot}
  \end{figure}
This section details our methodology for implementing a recent LLM family, Qwen3, on IMAX, our general-purpose CGLA accelerator. 
We constructed a hybrid execution model built upon the widely adopted llama.cpp framework, which offloads computationally intensive kernels to the IMAX accelerator. 
We first provide an overview of the overall execution framework, followed by a detailed description of the dataflow design and mapping strategies for the various quantized kernels we implemented in this work.

\subsection{Execution Models and Framework Overview}
For a realistic system-level evaluation, we implement a hybrid execution model based on the llama.cpp framework.
For our target models, we selected the state-of-the-art Qwen3. 
We selected the Qwen3 for its high performance at smaller scales and its wide range of model sizes (e.g., 0.6B, 1.7B, 8B). 
This allows us to test our accelerator's capabilities on workloads ranging from edge to high-performance scenarios.
In our model, the host CPU and the IMAX accelerator collaborate on LLM inference.
Fig.~\ref{fig:llama_arch} presents a high-level view of the target LLM architecture and delineates our proposed task partitioning.
This partitioning is guided by the principle of assigning tasks to the most suitable processing unit: utilizing the CPU for complex, sequential control flow and the IMAX accelerator for parallel, computationally intensive operations.
The host CPU handles tasks that demand complex control and sequential logic. 
These responsibilities include prompt tokenization, embedding layer computations, KV cache management, and the final Softmax operation. 
Conversely, we offload the most computationally demanding kernels, which constitute the majority of the inference latency, to the IMAX accelerator. 
As highlighted in pink in Fig.~\ref{fig:llama_arch}, IMAX executes the dot-product operations within all linear projections, the Grouped Multi-Query Attention mechanism, and the linear transformations of the SwiGLU network. 
We retain non-linear operations, such as RMS Normalization and the application of Rotary Position Encodings, on the host CPU. 
We designed this functional partitioning to maximize overall system throughput by utilizing the architectural strengths of each processing unit.

\subsection{Quantization and Supported Kernels}
Quantization is an essential technique in LLM inference. 
It reduces both the memory footprint and the required memory bandwidth, while also enabling faster computation through integer arithmetic. 
In resource-constrained environments, model compression via quantization is often a prerequisite for execution. 
The llama.cpp framework supports a diverse range of quantization schemes, offering flexibility to balance the trade-off between model accuracy and performance.

To demonstrate the versatility and robustness of the IMAX architecture, we implemented and evaluated the following four distinct computational kernels, each with varying characteristics:
\begin{itemize}
    \item \textbf{FP16}: A \SI[number-unit-product=\nobreakdash-]{16}{bit} floating-point format. This serves not only as a baseline but is also an essential data type used for specific high-precision operations within all quantized models.
    \item \textbf{Q8\_0}: A standard \SI[number-unit-product=\nobreakdash-]{8}{bit} integer quantization scheme. This kernel constitutes the majority of the operations performed in the Q8\_0 models.
    \item \textbf{Q3\_K}: A highly compressed mixed-precision scheme by combining \SI[number-unit-product=\nobreakdash-]{1}{bit}, \SI[number-unit-product=\nobreakdash-]{2}{bit}, \SI[number-unit-product=\nobreakdash-]{4}{bit}, and \SI[number-unit-product=\nobreakdash-]{6}{bit} quantization blocks. It represents the majority of the computations in the highly compressed Q3\_K\_S models.
    \item \textbf{Q6\_K}: A \SI[number-unit-product=\nobreakdash-]{6}{bit} integer format that is also utilized for specific layers within the Q3\_K\_S models, complementing the Q3\_K kernel.
\end{itemize}

These data types are strategically employed across different layers of the models. 
The large weight matrices of the linear layers contain the majority of model parameters in the attention and feed-forward networks. 
These matrices are quantized to low-bit integer formats such as Q8\_0, Q3\_K, and Q6\_K.
This approach substantially reduces the overall model file size and memory bandwidth requirements. 
In contrast, we preserve the weights of the normalization layers in high-precision FP16 to avoid the risk of performance degradation, as these layers are necessary to maintain computational stability.
Since the parameter count in normalization layers is negligible compared to that of linear layers, retaining their precision has a minimal impact on the total model size. 
Our evaluation adopts a common strategy used in modern LLMs. 
We quantize only the large linear layers, while the smaller normalization layers remain in high precision to maintain model stability.

The selection of these specific quantization formats is intended to cover a realistic spectrum of quality-efficiency trade-offs. 
Empirical studies on the Qwen3 model family have demonstrated that \SI[number-unit-product=\nobreakdash-]{8}{bit} quantization (Q8\_0) maintains performance nearly identical to the FP16 baseline, with negligible degradation on standard benchmarks such as MMLU\cite{qwen3_quantization}. 
Conversely, quantization formats with extremely low bit-widths, such as Q3\_K, result in a measurable degradation in accuracy.
However, their reduced memory footprint, a \num{4.5}$\times$ reduction compared to FP16, makes them an enabling technology for deploying LLMs on severely memory-constrained edge devices. 
Our evaluation includes this range to demonstrate the architecture's flexibility across this entire spectrum.

\subsection{Kernel Mapping on IMAX}


This section details how high-level dot-product operations from llama.cpp are compiled and mapped onto the IMAX CGLA architecture. 
The mapping process leverages IMAX's compiler-friendly design and a rich set of custom instructions to maximize pipelining and data-level parallelism.
First, the compiler analyzes the high-level C++ code, focusing on performance-critical loops such as the dot-product operation.
Then, it maps the dataflow of the operation onto IMAX's one-dimensional PE array. 
This linear topology, unlike 2D mesh architectures, allows for a deterministic mapping without complex routing heuristics, enabling predictable performance. 
Second, the compiler translates the low-level arithmetic within the dataflow into IMAX's custom instructions, which are designed to exploit fine-grained data parallelism as shown in Fig.~\ref{fig:q8dot}. 
The following subsections describe the specific dataflows and custom instructions used for each implemented kernel, serving as concrete examples of this mapping process.

Fig.~\ref{fig:f16dot} illustrates the dataflow for the FP16 dot-product kernel, which we designed to use the full programmability of IMAX. 
We first employ a lookup table (LUT) implemented within each PE to efficiently convert incoming FP16 data to an internal FP32 representation in-line, thus bypassing the need for dedicated conversion hardware. 
To enhance throughput, we then exploit two key parallelization features of IMAX. 
First, we apply SIMD instructions to execute two \SI[number-unit-product=\nobreakdash-]{32}{bit} operations concurrently on a single \SI[number-unit-product=\nobreakdash-]{64}{bit} datapath, maximizing the utilization of the Fused Multiply-Add (FMA) units. 
Second, we use column-wise multithreading to effectively mask the arithmetic pipeline latency. 
This technique time-multiplexes multiple logical FMA operations on a single physical FPU.
This kernel utilizes \num{22} arithmetic units to efficiently process a \num{16}-element multiplication in a single operational burst.

\begin{figure}[t]
  \centering
  \includegraphics[width=0.95\columnwidth]{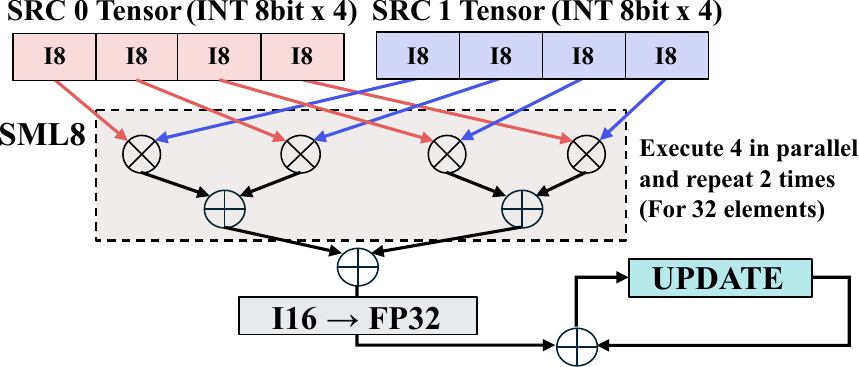}
  \caption{Dataflow for the Q8\_0 dot-product operation with one unit. This kernel handles the 8-bit integer format.}
  \label{fig:q8dot_one_unit}
\end{figure}

The dataflow for the Q8\_0 dot-product operation, shown in Fig.~\ref{fig:q8dot}, serves as the architectural foundation for all our quantized kernels. 
The IMAX dataflow efficiently integrates multiplication, addition, and data type conversion. 
We accelerate the core multiply-accumulate operation using the custom instructions conceptually illustrated in Fig.~\ref{fig:q8dot_one_unit}. 
The OP\_SML8 instruction performs a two-way SIMD signed \SI[number-unit-product=\nobreakdash-]{8}{bit} multiply-accumulate, independently multiplying each \SI[number-unit-product=\nobreakdash-]{8}{bit} segment of the input operands and summing the results into a sign-extended \SI[number-unit-product=\nobreakdash-]{24}{bit} output. 
We then use the OP\_AD24 instruction, a two-way \SI[number-unit-product=\nobreakdash-]{24}{bit} integer addition, to aggregate these intermediate results along the pipeline. 
Data is pipelined across twelve PEs and accumulated as \SI[number-unit-product=\nobreakdash-]{24}{bit} integers before being multiplied by a single-precision \SI[number-unit-product=\nobreakdash-]{32}{bit} floating-point scaling factor in the final stage. 
This entire process is replicated four times to run in parallel across the PE array, and two such parallel executions complete the processing of a full \num{32}\nobreakdash-element vector segment.
The Q8\_0 kernel utilizes a total of \num{46} arithmetic units.

\begin{figure}[t]
  \centering
  \includegraphics[width=0.95\columnwidth]{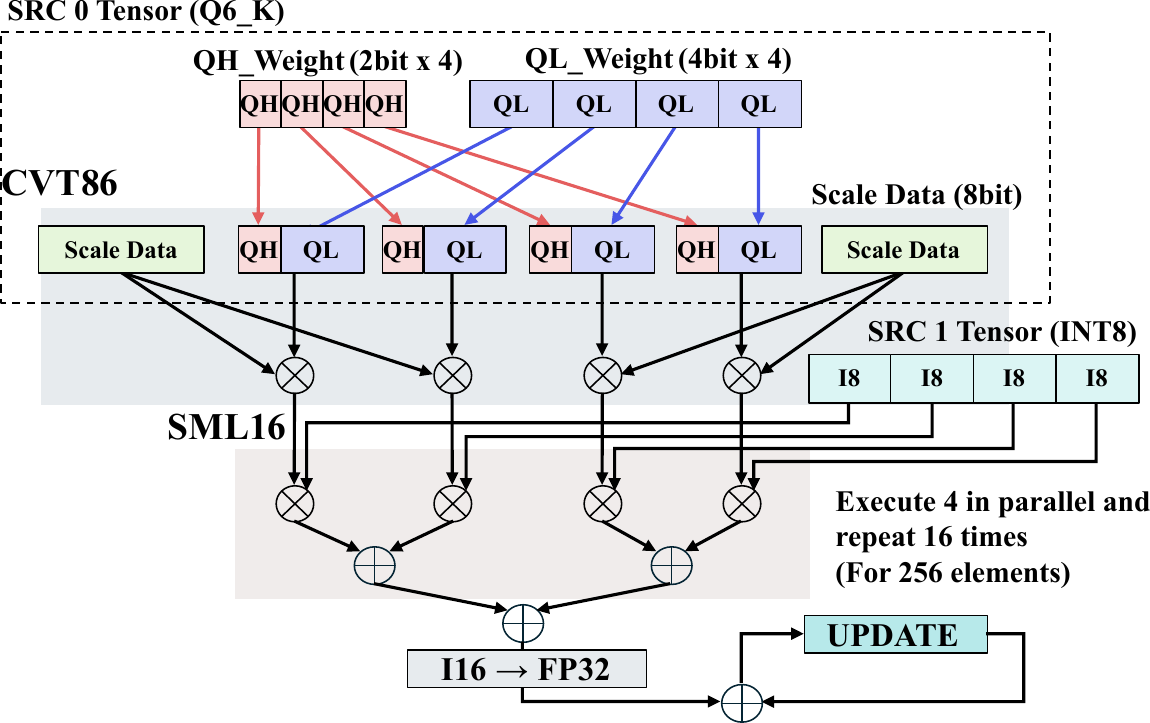}
  \caption{Dataflow for the Q6\_K dot-product operation. This kernel handles the Q6\_K format, which packs 2-bit (QH) and 4-bit (QL) quantized weights.}

  \label{fig:q6kdot}
\end{figure}

\begin{figure}[t]
    \centering
    \includegraphics[width=0.95\columnwidth]{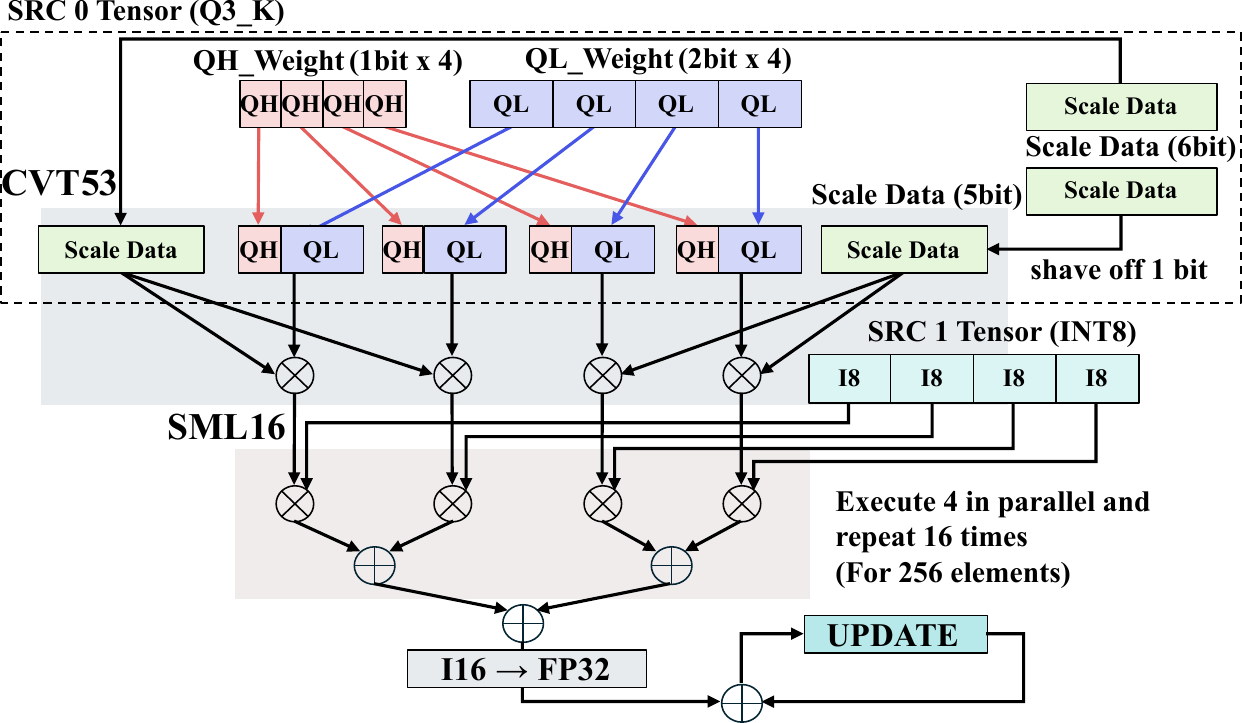}
    \caption{Dataflow for the Q3\_K dot-product operation. This kernel handles the Q3\_K format, which packs 1-bit (QH) and 2-bit (QL) quantized weights.}
    \label{fig:q3kdot}
  \end{figure}
  
The dataflows for the Q6\_K and Q3\_K dot-product kernels, shown in Fig.~\ref{fig:q6kdot} and Fig.~\ref{fig:q3kdot}, respectively, build upon this Q8\_0 foundation. 
Although these low-bit mixed-precision kernels involve more complex data structures, they adopt a unified processing flow to maintain architectural versatility. 
The core strategy is to decompress and reconfigure the diverse low-bit formats into a common \SI[number-unit-product=\nobreakdash-]{8}{bit} integer (INT8) representation at the front-end, allowing the standardized back-end multiply-accumulate pipeline to be reused without modification.
The Q6\_K kernel utilizes \num{64} arithmetic units. For this kernel, a custom CVT86 instruction decodes incoming \SI[number-unit-product=\nobreakdash-]{2}{bit} and \SI[number-unit-product=\nobreakdash-]{4}{bit} quantized weights and their corresponding \SI[number-unit-product=\nobreakdash-]{8}{bit} scales in a single cycle, producing \SI[number-unit-product=\nobreakdash-]{16}{bit} intermediate data. Another custom instruction, SML16, then performs the dot-product operation between this decoded data and the \SI[number-unit-product=\nobreakdash-]{8}{bit} integer inputs.
This sequential execution of decompression followed by computation is a methodical approach to managing complex, packed data formats within the CGLA architecture.
The Q3\_K kernel, which uses \num{51} arithmetic units, requires even more intricate data manipulation. 
This format natively uses \SI[number-unit-product=\nobreakdash-]{6}{bit} scales with \SI[number-unit-product=\nobreakdash-]{2}{bit} and \SI[number-unit-product=\nobreakdash-]{1}{bit} quantized weights. 
To process this data efficiently on IMAX's SIMD architecture, we designed a custom data reconfiguration method. 
A dedicated instruction, OP\_CVT53, performs an approximate conversion of the \SI[number-unit-product=\nobreakdash-]{6}{bit} scales to \SI[number-unit-product=\nobreakdash-]{5}{bit} and packs the \SI[number-unit-product=\nobreakdash-]{2}{bit} and \SI[number-unit-product=\nobreakdash-]{1}{bit} segments into a unified \SI[number-unit-product=\nobreakdash-]{3}{bit} format. 
This reconfiguration enables the combination of \SI[number-unit-product=\nobreakdash-]{8}{bit} input data with \SI[number-unit-product=\nobreakdash-]{5}{bit} and \SI[number-unit-product=\nobreakdash-]{3}{bit} weight data, allowing us to implement a processing flow similar to that of the Q8\_0 kernel. 
We empirically confirmed that this approximation of the scale data, which carries less information than the weights, has a negligible impact on the final computational accuracy. 
This front-end conversion approach allows us to achieve high performance without architectural versatility, processing \num{256} elements per burst by running four parallel dataflows for sixteen iterations.

\subsection{LMM Management and Configuration}

\begin{table*}[t]
  \centering
  \begin{threeparttable}
  
  \caption{Physical specifications and performance comparisons of various devices.}
  \label{tab:processor_comparison_annotated}
  
  \begin{tabular}{@{} llrrrrrrr l @{}} 
    \toprule
    \textbf{Device} & \textbf{CPU} & \textbf{Cores} & \textbf{Chip area} & \textbf{Process node} & \textbf{Frequency} & \textbf{Memory} & \textbf{Power}\tnote{c} \\
    & & & \textbf{(mm$^2$)} & \textbf{(nm)} & \textbf{(MHz)} & & \textbf{(W)} \\
    \midrule
 
    \textbf{IMAX3 (Xilinx VPK180)} & Arm Cortex-A72 & 64\tnote{a} & - & 7 & 145 & \SI{8}{\giga\byte} + \SI{4}{\giga\byte} DDR4\tnote{b} & 180 \\
 
    \textbf{IMAX3 (\SI{28}{\nano\meter})} & - & 64\tnote{a} & 14.6 & 28 & 840 & - & 2.16 - 6.1 \\

    \textbf{NVIDIA RTX 4090} & Xeon W5-2455X & 16384 & 608 & 5 & 2520 & \SI{24}{\giga\byte} + \SI{4}{\giga\byte} DDR6 & 450 \\
    \textbf{NVIDIA GTX 1080 Ti} & Xeon W5-2455X & 3584 & 448 & 16 & 1582 & \SI{11}{\giga\byte} DDR5 & 250 \\
    \textbf{Jetson AGX Orin 32GB} & Arm Cortex-A78AE & 1792 & 200 & 8 & 930 & \SI{32}{\giga\byte} DDR5 & 60 \\
    \bottomrule
  \end{tabular}

 \begin{tablenotes}[para,flushleft]
   \item[a] The number of cores for IMAX3 refers to the number of PEs per lane. 
   \item[b] \SI{8}{\giga\byte} DDR4 for OS buffer and \SI{4}{\giga\byte} DDR4 for DMA buffer.
  \item[c] For IMAX3~(\SI{28}{\nano\meter}), the power consumption is estimated and varies depending on the kernels (FP16: \SI{2.16}{\watt}, Q8\_0: \SI{4.41}{\watt}, Q3\_K: \SI{4.88}{\watt}, Q6\_K: \SI{6.1}{\watt}), references for other devices are from Cortex-A72\nobreak\cite{Versal}, Jetson AGX Orin \SI{32}{\giga\byte} (most high performance mode)\nobreak\cite{nvidia_jetson_agx_orin}, NVIDIA GTX 1080 Ti\nobreak\cite{nvidia_1080ti}, NVIDIA RTX 4090\nobreak\cite{nvidia_4090}.
 \end{tablenotes}
  
   \end{threeparttable}
  \end{table*}

Although the LMMs are configurable up to \SI{512}{K\byte}, we selected a size of \SI{64}{K\byte} for all evaluations in this work. 
We found this configuration to be a favorable compromise between power consumption and performance, as it is sufficient to accommodate the tensor sizes involved in the dot-product operations of the Qwen3 models we evaluated. 
We will provide a quantitative justification for this choice in the discussion in Section~\ref{discussion}.
Data transfers between the host and IMAX via DMA can become a significant performance bottleneck, particularly for large-scale models. 
To mitigate this, we implemented a transfer coalescing strategy to maximize DMA efficiency. 
A naive implementation would issue separate DMA transactions for each input tensor (e.g., activations, weights, and scaling factors), incurring substantial overhead from multiple transaction setups. 
Although these tensors may reside in non-contiguous memory locations, our approach aggregates them into a single, contiguous block in the host-side DMA buffer before initiating the transfer.
For instance, the Q8\_0 kernel requires four distinct input arrays. 
By arranging them contiguously, the IMAX DMA engine, which utilizes a shared address space, can load the entire data block into the LMMs with a single burst-transfer instruction. 
A similar coalescing strategy is applied when writing results back to the host. 
This method significantly improves data transfer efficiency by minimizing the overhead associated with issuing multiple DMA transactions.
A preliminary evaluation of this optimization confirmed its effectiveness, accelerating the LOAD phase by a factor of 1.2 and the DRAIN phase by a factor of 4.8 compared to the naive implementation, which is a non-coalesced approach. 
All results reported in Section~\ref{ex_and_re} utilize this optimized data transfer method, demonstrating the critical importance of such low-level optimizations in memory-bound LLM workloads.

%% file: experiments_and_results.tex
\section{Experiments and Results}
\label{ex_and_re}
In this section, we present a comprehensive, E2E evaluation of the LLM inference performance on our IMAX accelerator. 
We compare the measured performance of our FPGA prototype and the projected performance of a potential \SI{28}{\nano\meter} ASIC implementation against a state-of-the-art high-performance GPU, a general-purpose GPU with a comparable process node, and a dedicated edge AI GPU. 
The primary objective of these experiments is to quantitatively evaluate the performance and energy efficiency of the IMAX architecture. 
This analysis focuses on the fundamental trade-off between performance and power consumption.

\subsection{Experimental Setup}
\begin{figure}[t]
  \centering
  \includegraphics[width=1\columnwidth]{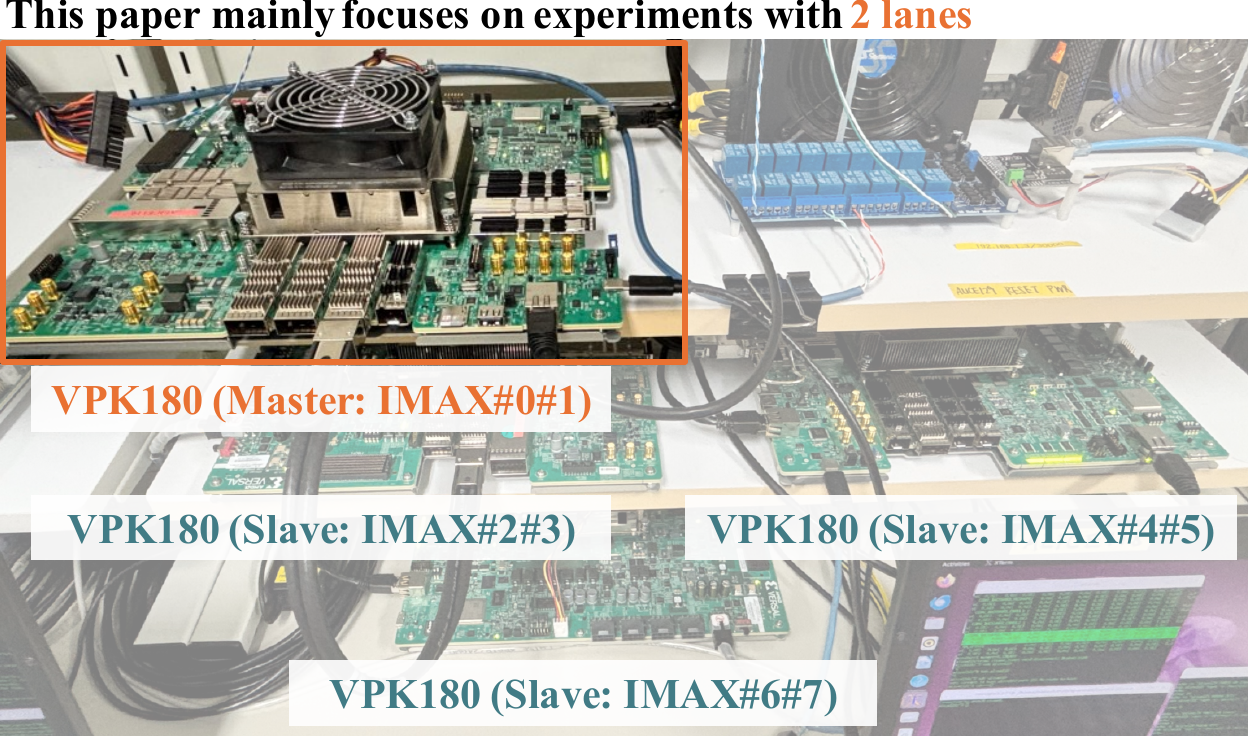}
  \caption{The experimental setup for the IMAX3 FPGA prototype, featuring a multi-board configuration with four AMD Versal VPK180 evaluation kits.}
  \label{fig:imax3_proto}
\end{figure}

We compare the performance of our IMAX accelerator against several commercial platforms. 
As shown in Fig.~\ref{fig:imax3_proto}, we implemented the IMAX prototype on an AMD Versal Premium VPK180 evaluation kit using Vivado 2024.1. 
This prototype consists of an eight-lane IMAX operating at \SI{145}{\mega\hertz} and uses an Arm Cortex-A72 PS as the host.
Although the FPGA supports up to eight IMAX lanes, our main evaluation uses a two-lane configuration. 
A preliminary analysis revealed that the dual-core ARM host becomes a performance bottleneck beyond two lanes, limiting the accelerator's utilization. 
This two-lane setup was therefore chosen to isolate the accelerator's performance from the host system. 
The architecture's scalability and the impact of host performance are further analyzed in Section~\ref{discussion}.
To evaluate the future potential of the architecture, we projected its performance as a \SI{28}{\nano\meter} ASIC. 
This projection is based on the established methodology from our previous work\nobreak\cite{imax3}, where we performed a detailed static timing and power analysis using Synopsys Design Compiler\cite{SynopsysNDDCUltra}. 
The analysis utilized the TSMC \SI{28}{\nano\meter} process technology, including its standard cell libraries and wireload models, and assumed \SI{10}{\percent} average switching activity for power estimation. 
From this analysis, we determined that a maximum operating frequency of \SI{840}{\mega\hertz} is achievable, which represents an approximately \num{6}$\times$ speedup over the FPGA implementation.
For the \SI{64}{K\byte} LMM configuration adopted in our evaluation, the power was calculated to be \SI{2.16}{\watt} for the FP16 kernel, \SI{4.41}{\watt} for Q8\_0, \SI{4.88}{\watt} for Q3\_K, and \SI{6.1}{\watt} for Q6\_K. 
Our commercial comparison platforms include a modern high-performance GPU system (NVIDIA RTX 4090), a previous-generation GPU system (NVIDIA GTX 1080 Ti), and an edge AI device (NVIDIA Jetson AGX Orin). 
The GTX 1080 Ti was selected to provide a comparison against a widely adopted GPU manufactured on a \SI{16}{\nano\meter} process, which is closer to our \SI{28}{\nano\meter} ASIC projection. 
Detailed hardware specifications are summarized in Table~\ref{tab:processor_comparison_annotated}.
The software and operating system configurations for each platform were standardized to ensure a fair comparison:
\begin{itemize}
\item The IMAX FPGA prototype runs a PetaLinux distribution, cross-compiled for the 64-bit ARM (aarch64) architecture.
\item Both the NVIDIA RTX 4090 and GTX 1080 Ti systems operate on Ubuntu 22.04.5 LTS with CUDA Toolkit 12.4.
\item The NVIDIA Jetson AGX Orin uses the JetPack R36.4.4 software stack.
\end{itemize}
All platforms were benchmarked using an identical version of the llama.cpp framework and the exact same quantized model files. 
This rigorous setup eliminates variations from the software stack and model data, allowing for a direct comparison of the underlying hardware architectures.

To establish a consistent baseline for comparing energy efficiency, our power model utilizes nominal specifications.
For the IMAX (\SI{28}{\nano\meter}) ASIC projection, the total power during offloaded phases is calculated based on synthesis results. 
Specifically, the active power is determined by multiplying the power estimated from synthesis by the number of active lanes (two in our primary evaluation). 
The total system power is then calculated by adding the measured idle power of the host CPU to this active power value. 
This model distinguishes between host-primary processing and phases where the IMAX cores are active.
For the commercial GPU/CPU platforms, we model active power using their official Thermal Design Power (TDP) values applying either the host CPU's base or peak TDP\cite{Intel_Xeon_w5-2455X_Specs} depending on which component is primarily active. 
This approach enables us to assess the potential energy efficiency of each architecture operating within its specified maximum power budget.
While TDP does not represent average power consumption, it provides a standardized metric for comparing performance under peak load conditions. 
The NVIDIA Jetson AGX Orin was evaluated in its nominal \SI{60}{\watt} maximum performance mode. 
We acknowledge this methodological choice as a limitation and discuss its implications in Section~\ref{discussion}.

We executed a diverse set of LLM inference workloads on these platforms.
We selected three models with varying parameter counts (0.6B, 1.7B, and 8B) from the state-of-the-art Qwen3 LLM family. 
We then combined these models with the multiple quantized kernels implemented in Section~\ref{proposed}, resulting in 54 distinct workloads to test the architecture's versatility and robustness. 
For these experiments, we varied the combination of input and output tokens from [8:1] to [32:16]. 
This range is intended to cover various practical scenarios. 
For instance, short input-output pairs mimic latency-sensitive Q\&A in conversational AI, whereas long inputs with short outputs represent tasks such as document summarization.
Combinations requiring longer outputs correspond to throughput-dependent scenarios such as code generation or lengthy text translation.
All reported performance metrics (E2E latency, PDP, and EDP) are the average of 10 independent runs for each workload configuration.
A fixed seed was used for all experiments to ensure reproducibility.
The standard deviation for all measurements was consistently below \SI{3}{\percent} of the mean value, indicating high stability and minimal run-to-run variation.

We assess the performance of each platform using three primary metrics. 
The first metric is E2E latency, defined as the total time from prompt input to the generation of the first token, which indicates system responsiveness. 
The second is the PDP, which, as shown in~\eqref{eq:pdp}, directly measures energy efficiency as the total energy consumed to complete a task.
\begin{equation}
  \text{PDP} = \text{Latency} \times \text{Power}
  \label{eq:pdp}
\end{equation}
The third metric is the EDP, calculated as the product of power and the square of the latency~\eqref{eq:edp}.
The EDP offers a comprehensive view of the performance-energy trade-off.
\begin{equation}
  \text{EDP} =  \text{Latency}^2 \times \text{Power}
  \label{eq:edp}
\end{equation}
EDP is a key metric for power-constrained systems. 
We note that our PDP and EDP calculations are based on the nominal TDP values. 
Our analysis uses TDP, which reflects peak rather than average power consumption. 
Therefore, the results indicate the potential energy efficiency of each architecture, not its performance in typical application scenarios.

\begin{figure}[t]
  \centering
  
  \begin{subfigure}{1.0\columnwidth}
    \centering
    \includegraphics[width=\textwidth]{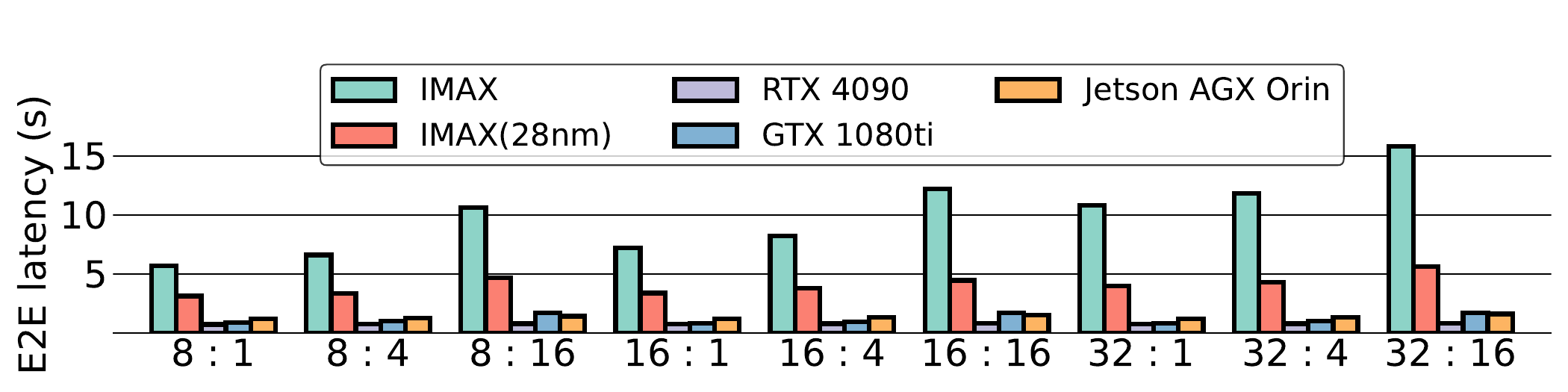}
    \subcaption{Qwen3-0.6B Q3\_K\_S}
    \label{fig:e2e_Qwen3_6B_Q3K_S}
  \end{subfigure}
  
  \vspace{1pt} 
  
  \begin{subfigure}{1.0\columnwidth}
    \centering
    \includegraphics[width=\textwidth]{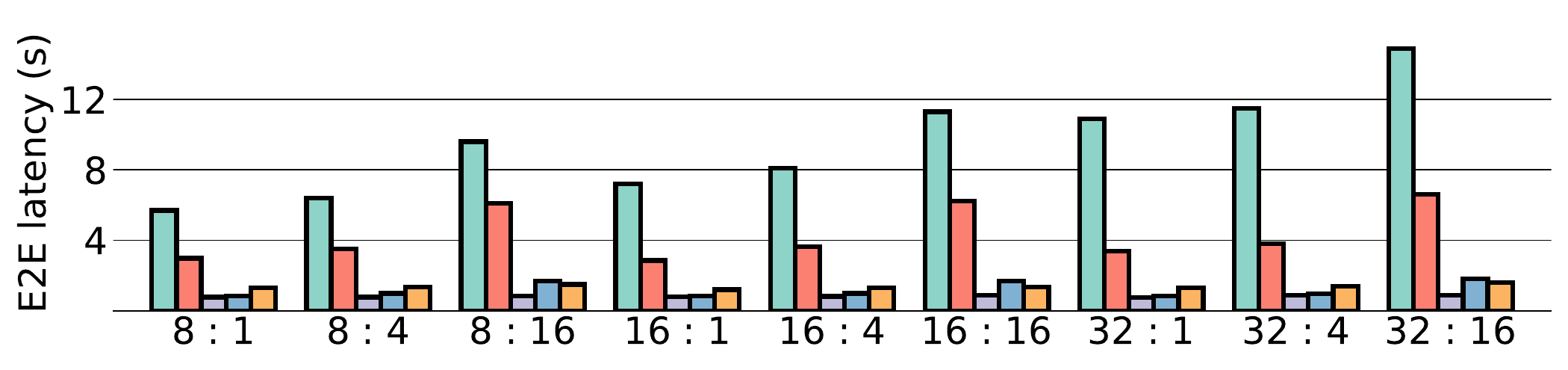}
    \subcaption{Qwen3-0.6B Q8\_0}
    \label{fig:e2e_Qwen3_6B_Q8_0}
  \end{subfigure}
  
  \vspace{1pt} 

    \begin{subfigure}{1.0\columnwidth}
      \centering
      \includegraphics[width=\textwidth]{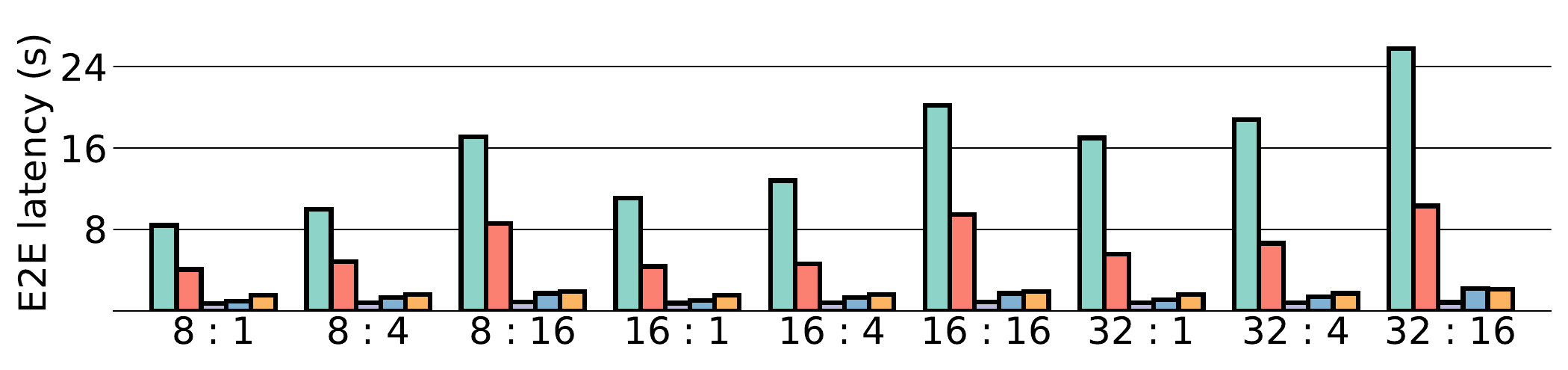}
      \subcaption{Qwen3-1.7B Q3\_K\_S}
      \label{fig:e2e_Qwen3_7B_Q3K_S}
    \end{subfigure}
    
    \vspace{1pt} 
    
    \begin{subfigure}{1.0\columnwidth}
      \centering
      \includegraphics[width=\textwidth]{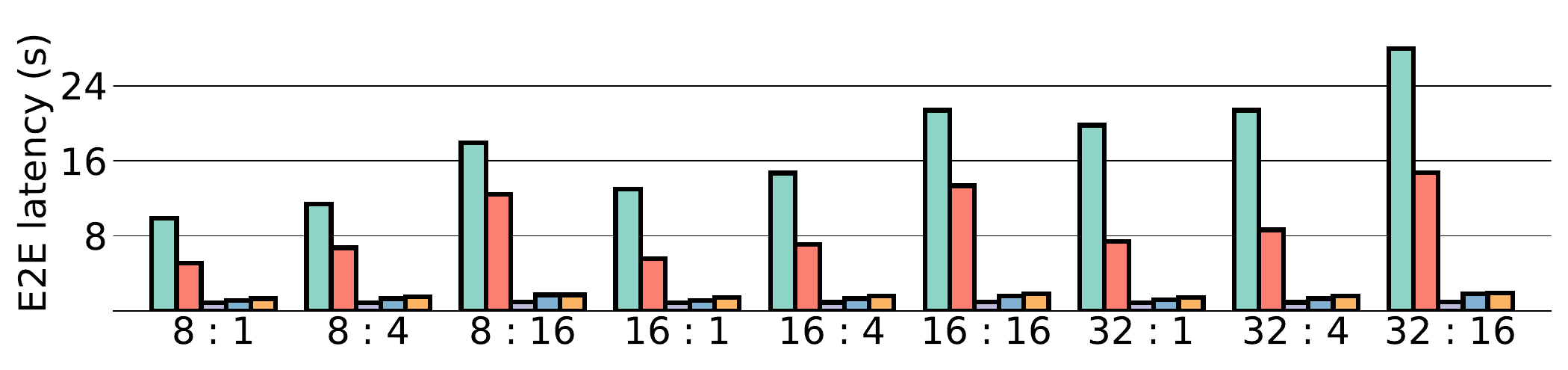}
      \subcaption{Qwen3-1.7B Q8\_0}
      \label{fig:e2e_Qwen3_7B_Q8_0}
    \end{subfigure}
    
    \vspace{1pt} 
  
  \begin{subfigure}{1.0\columnwidth}
    \centering
    \includegraphics[width=\textwidth]{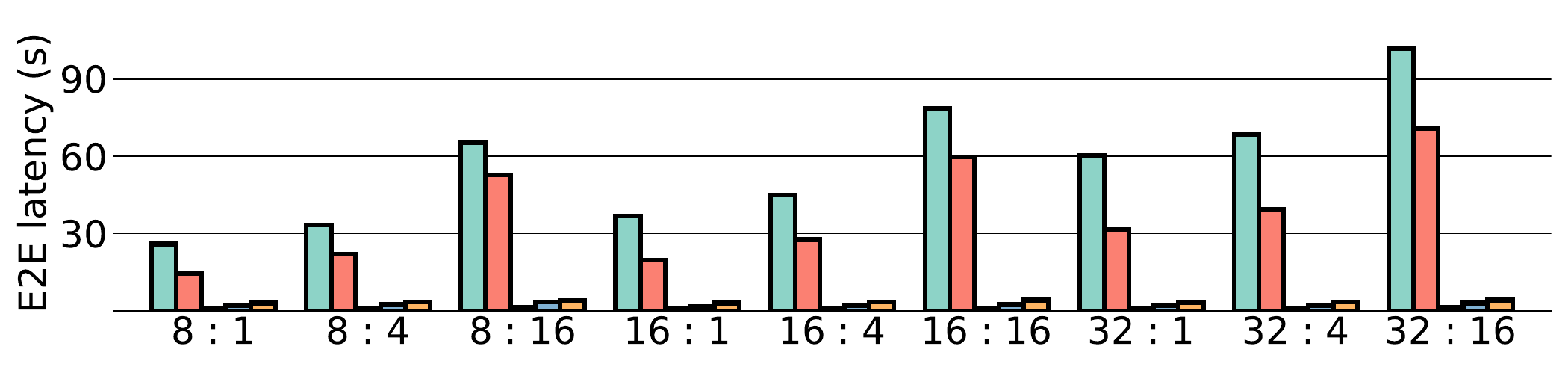}
    \subcaption{Qwen3-8B Q3\_K\_S}
    \label{fig:e2e_Qwen3_8B_Q3K_S}
  \end{subfigure}
  
  \vspace{1pt} 
  
  \begin{subfigure}{1.0\columnwidth}
    \centering
    \includegraphics[width=\textwidth]{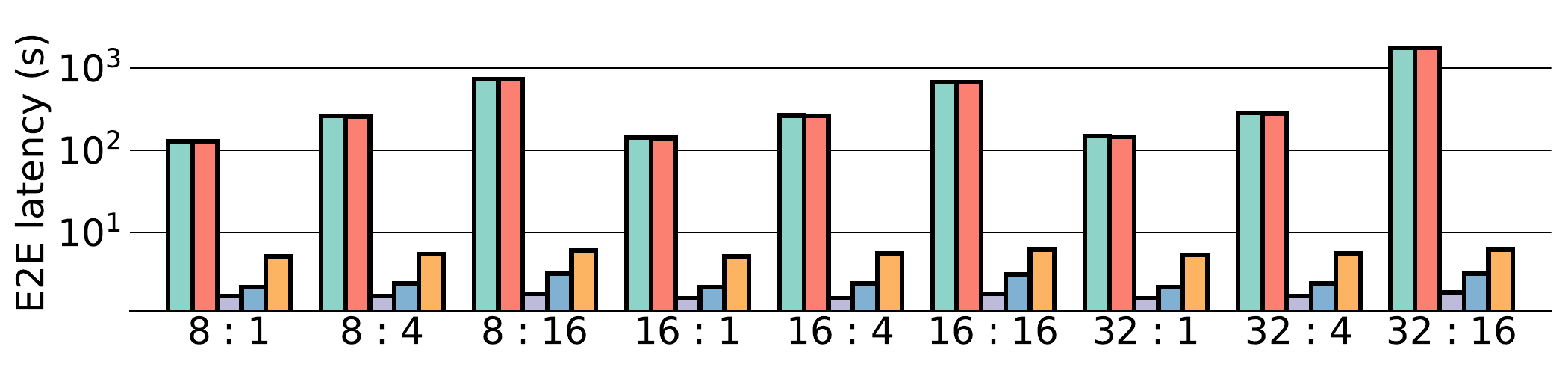}
    \subcaption{Qwen3-8B Q8\_0}
    \label{fig:e2e_Qwen3_8B_Q8_0}
  \end{subfigure}
  
  \caption{End-to-end performance comparison of E2E latency by device. IMAX is measured in FPGA and IMAX (\SI{28}{\nano\meter}) is projected.}
  \label{fig:e2e_latency}
\end{figure}

\begin{figure}[t]
  \centering

  \begin{subfigure}{1.0\columnwidth}
    \centering
    \includegraphics[width=\textwidth]{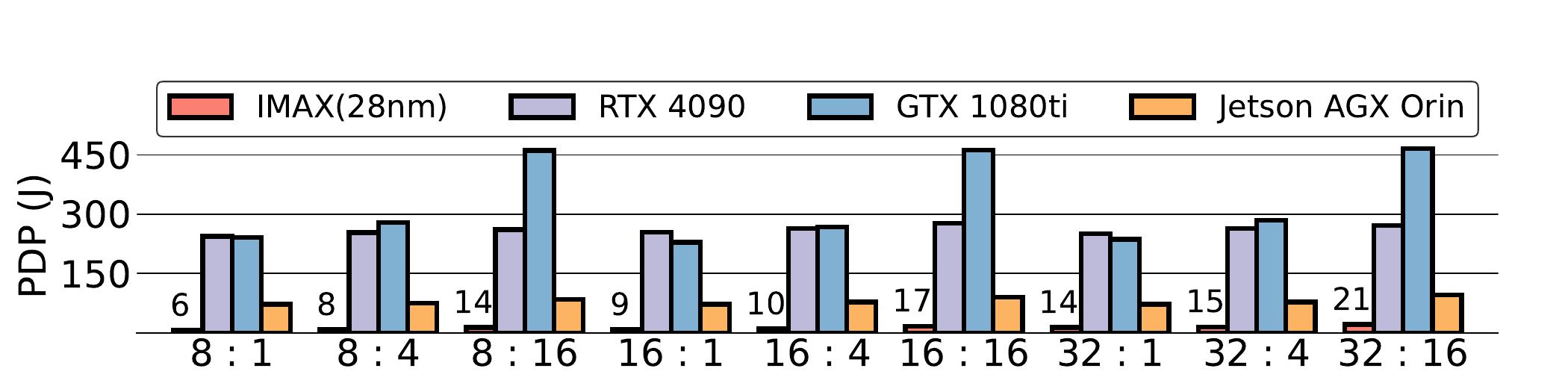}
    \subcaption{Qwen3-0.6B Q3\_K\_S}
    \label{fig:pdp_Qwen3_6B_Q3K_S}
  \end{subfigure}

  \vspace{1pt} 

  \begin{subfigure}{1.0\columnwidth}
    \centering
    \includegraphics[width=\textwidth]{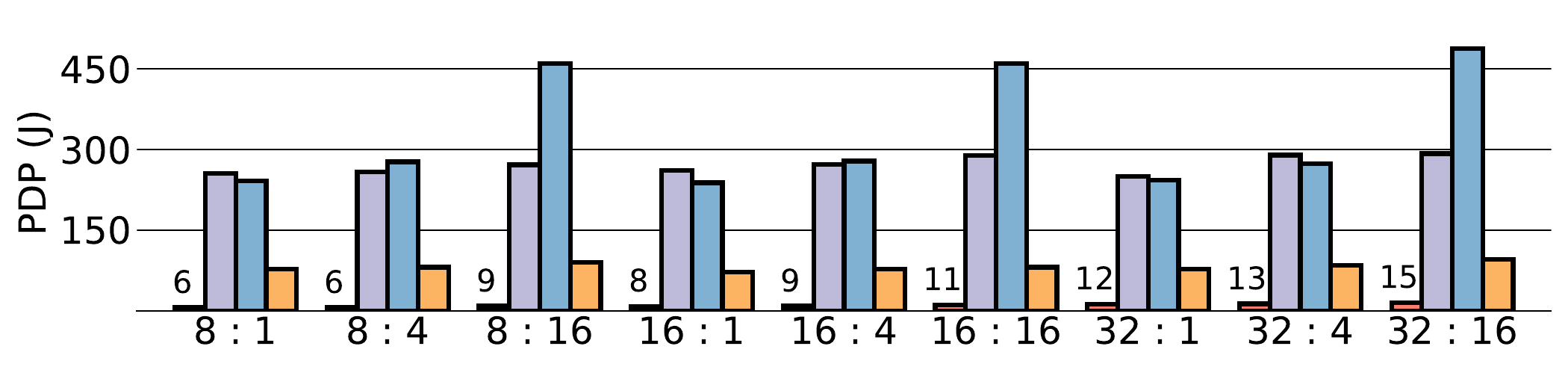}
    \subcaption{Qwen3-0.6B Q8\_0}
    \label{fig:pdp_Qwen3_6B_Q8_0}
  \end{subfigure}

  \begin{subfigure}{1.0\columnwidth}
    \centering
    \includegraphics[width=\textwidth]{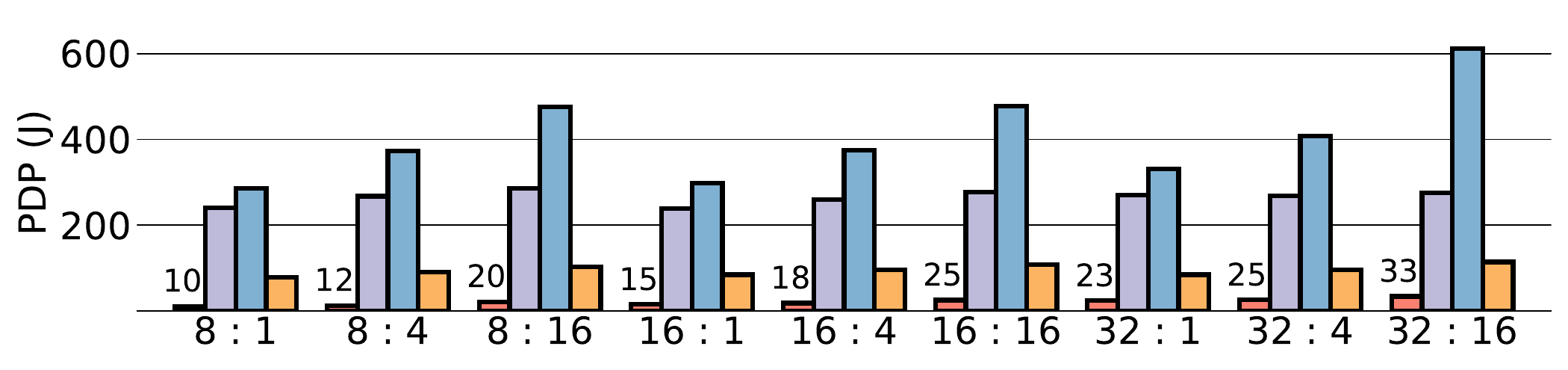}
    \subcaption{Qwen3-1.7B Q3\_K\_S}
    \label{fig:pdp_Qwen3_7B_Q3K_S}
  \end{subfigure}

  \vspace{1pt} 

  \begin{subfigure}{1.0\columnwidth}
    \centering
    \includegraphics[width=\textwidth]{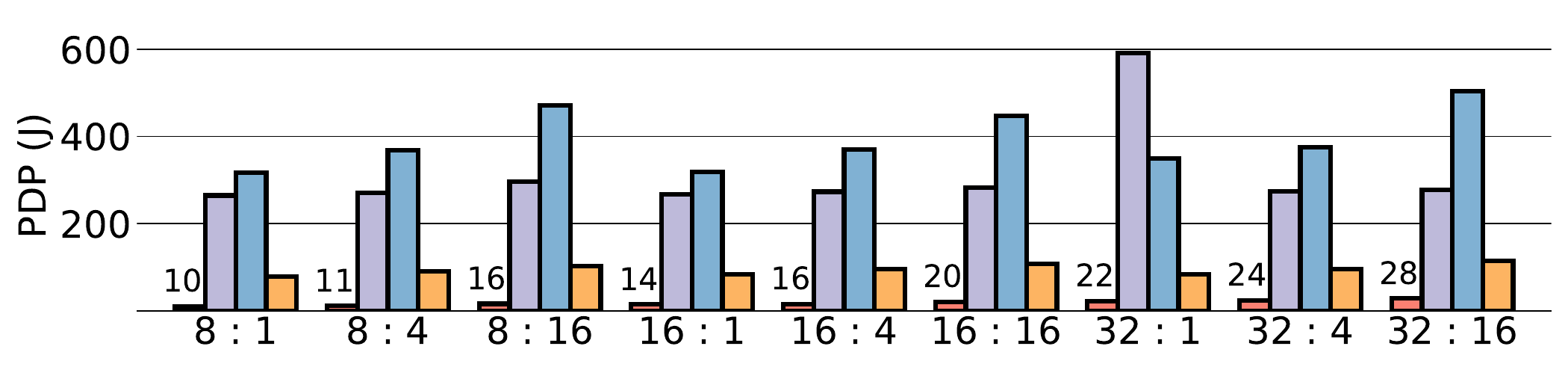}
    \subcaption{Qwen3-1.7B Q8\_0}
    \label{fig:pdp_Qwen3_7B_Q8_0}
  \end{subfigure}

  \vspace{1pt} 

  \begin{subfigure}{1.0\columnwidth}
    \centering
    \includegraphics[width=\textwidth]{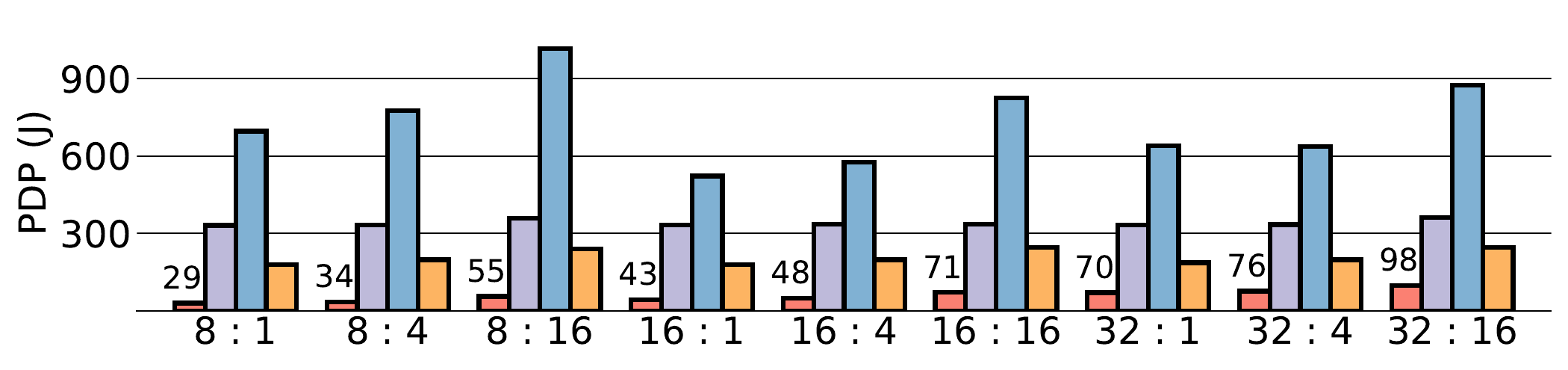}
    \subcaption{Qwen3-8B Q3\_K\_S}
    \label{fig:pdp_Qwen3_8B_Q3K_S}
  \end{subfigure}

  \vspace{1pt} 

  \begin{subfigure}{1.0\columnwidth}
    \centering
    \includegraphics[width=\textwidth]{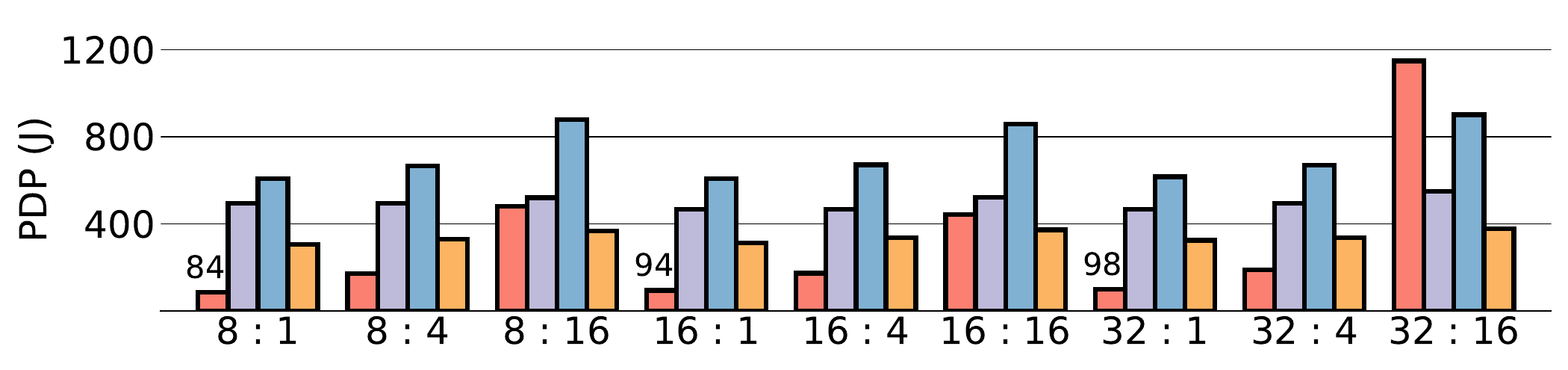}
    \subcaption{Qwen3-8B Q8\_0}
    \label{fig:pdp_Qwen3_8B_Q8_0}
  \end{subfigure}
  \caption{PDP comparison by device (lower is better). The numeric value for IMAX (\SI{28}{\nano\meter}) is shown when it is below \SI{100} to ensure readability.}
  \label{fig:e2e_pdp_comparison}
\end{figure}

\begin{figure}[t]
  \centering

  \begin{subfigure}{1.0\columnwidth}
    \centering
    \includegraphics[width=\textwidth]{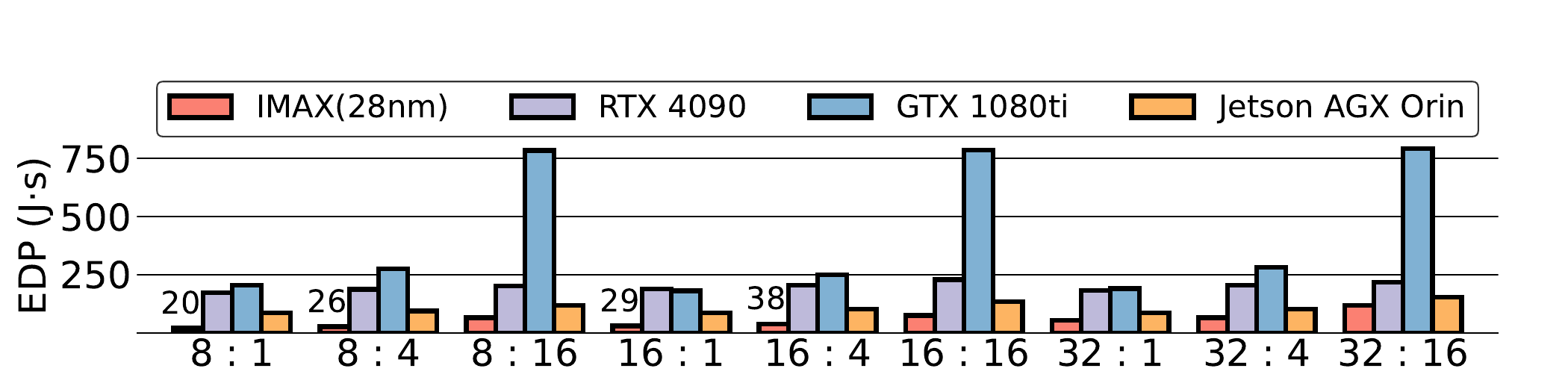}
    \subcaption{Qwen3-0.6B Q3\_K\_S}
    \label{fig:edp_Qwen3_6B_Q3K_S}
  \end{subfigure}

  \vspace{1pt} 

  \begin{subfigure}{1.0\columnwidth}
    \centering
    \includegraphics[width=\textwidth]{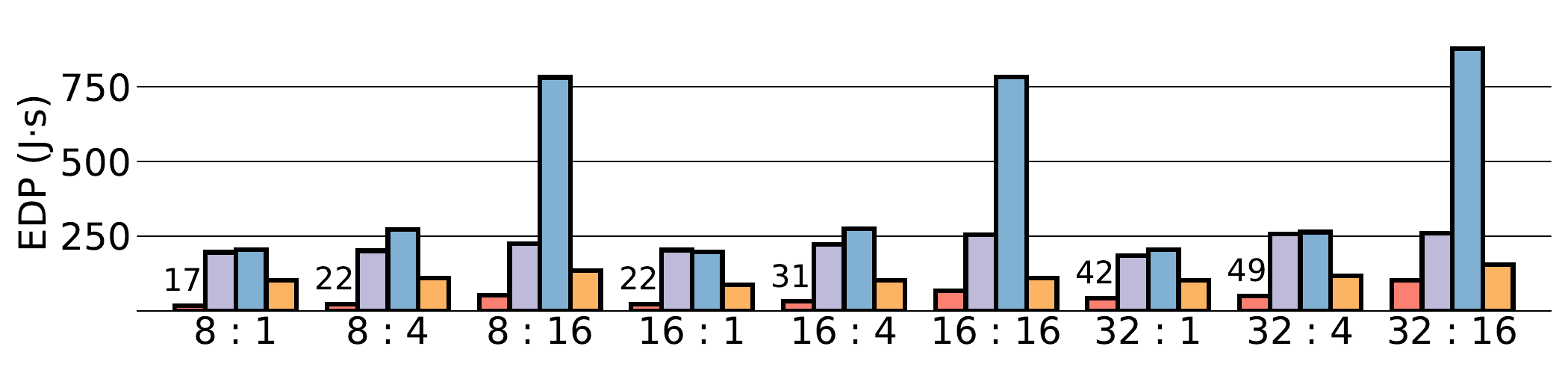}
    \subcaption{Qwen3-0.6B Q8\_0}
    \label{fig:edp_Qwen3_6B_Q8_0}
  \end{subfigure}

  \begin{subfigure}{1.0\columnwidth}
    \centering
    \includegraphics[width=\textwidth]{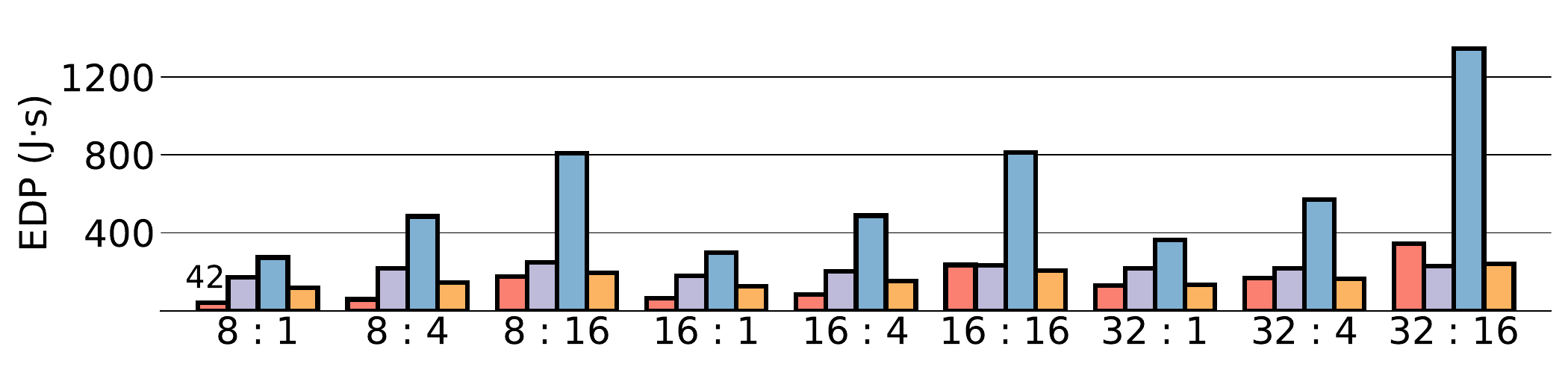}
    \subcaption{Qwen3-1.7B Q3\_K\_S}
    \label{fig:edp_Qwen3_7B_Q3K_S}
  \end{subfigure}

  \vspace{1pt} 

  \begin{subfigure}{1.0\columnwidth}
    \centering
    \includegraphics[width=\textwidth]{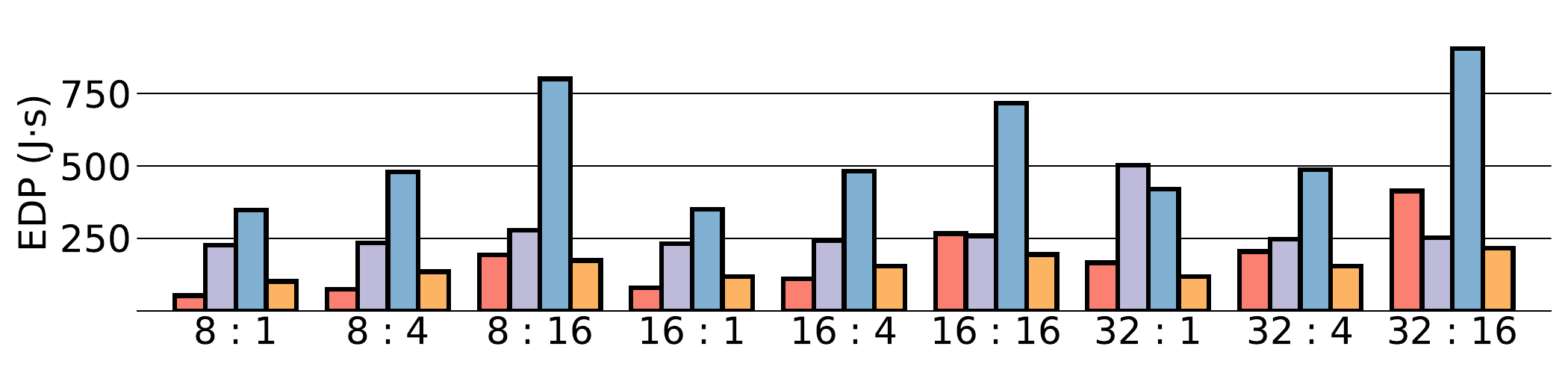}
    \subcaption{Qwen3-1.7B Q8\_0}
    \label{fig:edp_Qwen3_7B_Q8_0}
  \end{subfigure}

  \vspace{1pt} 

  \begin{subfigure}{1.0\columnwidth}
    \centering
    \includegraphics[width=\textwidth]{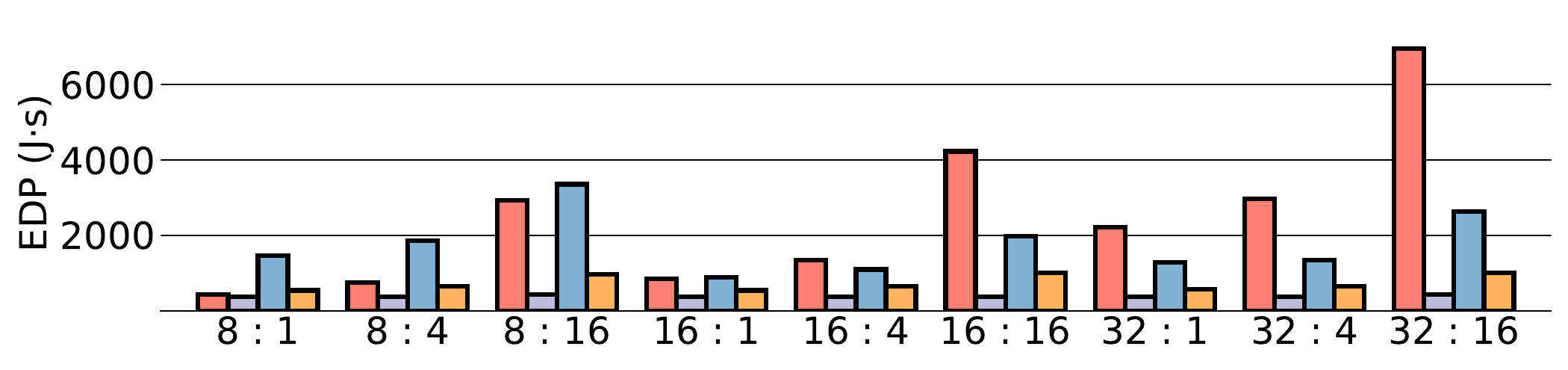}
    \subcaption{Qwen3-8B Q3\_K\_S}
    \label{fig:edp_Qwen3_8B_Q3K_S}
  \end{subfigure}

  \vspace{1pt} 

  \begin{subfigure}{1.0\columnwidth}
    \centering
    \includegraphics[width=\textwidth]{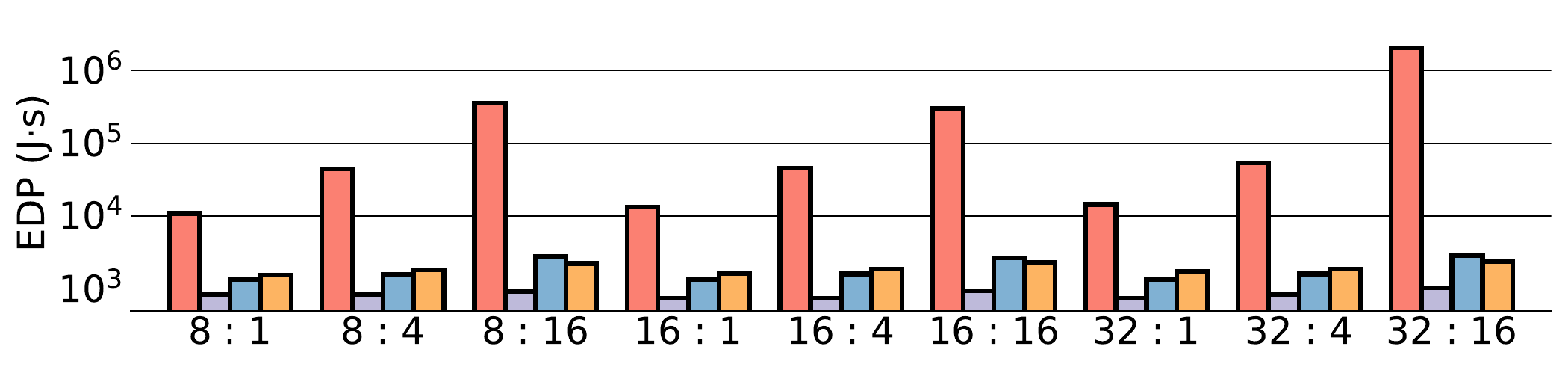}
    \subcaption{Qwen3-8B Q8\_0}
    \label{fig:edp_Qwen3_8B_Q8_0}
  \end{subfigure}
  \caption{EDP comparison by device (lower is better). The numeric value for IMAX (\SI{28}{\nano\meter}) is shown when it is below \SI{50} to ensure readability.}
  \label{fig:e2e_edp_comparison}
  
\end{figure}

\subsection{E2E Performance Evaluation}
In this subsection, we conduct an E2E performance evaluation of the IMAX architecture. 
We first analyze the E2E latency to assess pure processing speed, then shift our focus to energy efficiency, the primary emphasis of this work, by quantifying the PDP and EDP.

E2E latency comparison in Fig.~\ref{fig:e2e_latency} shows that data transfer has become the primary system bottleneck for our architecture. 
As expected, the NVIDIA RTX 4090 demonstrated the lowest latency in all scenarios due to its substantial resources. 
For instance, on a representative workload, the RTX 4090 achieved a latency of approximately \SI{0.8}{\second}, whereas our projected IMAX (\SI{28}{\nano\meter}) latency was \SI{5.63}{\second}.
Critically, this performance gap scales with model size, particularly for memory-bound models such as Qwen3-8B Q8\_0. 
This scaling behavior strongly indicates that the system's performance is not compute-bound by the IMAX core but is instead bottlenecked by data transfer overhead.

As shown in Fig.~\ref{fig:e2e_pdp_comparison} and Fig.~\ref{fig:e2e_edp_comparison}, the advantages of our design become clear from the energy-centric metrics.
These results indicate that the IMAX architecture, particularly the \SI{28}{\nano\meter} ASIC projection, has the potential to achieve energy efficiency far superior to existing platforms. 
In terms of PDP, the IMAX (\SI{28}{\nano\meter}) projection demonstrates significant advantages. 
For instance, on the compute-bound Qwen3-1.7B Q8\_0 [16:4] workload, IMAX achieved a PDP of just \SI{15.5}{\joule}, outperforming the RTX 4090 (\SI{28.4}{\joule}), GTX 1080 Ti (\SI{35.1}{\joule}), and Jetson (\SI{22.1}{\joule}). 
This represents an efficiency improvement of up to \num{44.4}$\times$, \num{54}$\times$, and \num{13.6}$\times$ over the respective platforms in certain workloads. 
However, this advantage diminishes as data transfer becomes the bottleneck. 
In the most memory-intensive case (Qwen3-8B Q8\_0 [32:16]), the PDP of IMAX surged to \SI{1148.7}{\joule}, which is substantially higher than that of the RTX 4090 (\SI{547.9}{\joule}) and the Jetson (\SI{378.0}{\joule}), highlighting the impact of data transfer overhead on energy consumption.

The EDP evaluation, which squares the impact of execution time, exposes the latency trade-offs of the architecture. 
In compute-bound workloads, the advantage of IMAX (\SI{28}{\nano\meter}) is particularly significant.
For example, on the Qwen3-0.6B Q3\_K\_S [32:16] workload, IMAX (\SI{28}{\nano\meter}) recorded an EDP of \SI{118.9}{\joule\second}, outperforming both the RTX 4090 (\SI{216.8}{\joule\second}) and the Jetson AGX Orin (\SI{153.6}{\joule\second}). 
It maintained a substantial efficiency lead in many scenarios over high-power GPUs, outperforming the RTX 4090 by up to \num{11.5}$\times$ and the GTX 1080 Ti by \num{15}$\times$.
However, as workloads become more memory-bound with larger data transfers, the impact of latency becomes the primary factor.
For the Qwen3-1.7B Q8\_0 [32:16] workload, the shorter latency of the Jetson (\SI{1.9}{\second}) gave it an advantage in EDP, resulting in a score of \SI{216.6}{\joule\second}\, that surpassed IMAX (\SI{413.6}{\joule\second}) with \SI{14.7}{\second}\ latency.
This result indicates the trade-off where low latency is prioritized in EDP, even though IMAX held the advantage in pure energy efficiency (PDP). 
This trend was most significant in the memory-bottlenecked Qwen3-8B Q8\_0 model, where the EDP of IMAX (\SI{28}{\nano\meter}) was substantially higher than that of the other platforms. 
These results suggest that, under the current system configuration, the EDP advantages of the IMAX (\SI{28}{\nano\meter}) architecture are most apparent in compute-bound workloads.

%% file: discussion.tex
\section{Discussion}
\label{discussion}

While Section~\ref{ex_and_re}  demonstrated the significant energy efficiency advantages of the CGLA-based approach, it also highlighted performance bottlenecks under specific conditions. 
To provide architectural insight and guide future work, this section analyzes these limitations from three perspectives: the impact of internal memory size, a granular breakdown of execution phase timings, and the inherent scalability limits imposed by the host system.
\subsection{Impact of LMM Size}
The size of the LMM in each PE is a key architectural design parameter.
This choice directly impacts overall system efficiency, as the LMM capacity governs the fundamental trade-off between performance and power consumption. 
In this subsection, we quantitatively validate our selection of a size of \SI{64}{K\byte} for the LMM, demonstrating that it provides a suitable balance between offload ratio and power efficiency for our target workloads.

\begin{figure}[t]
  \centering
  \subfloat[Qwen3-0.6B Q3\_K\_S\label{fig:sub_a}]{%
    \includegraphics[width=0.23\textwidth]{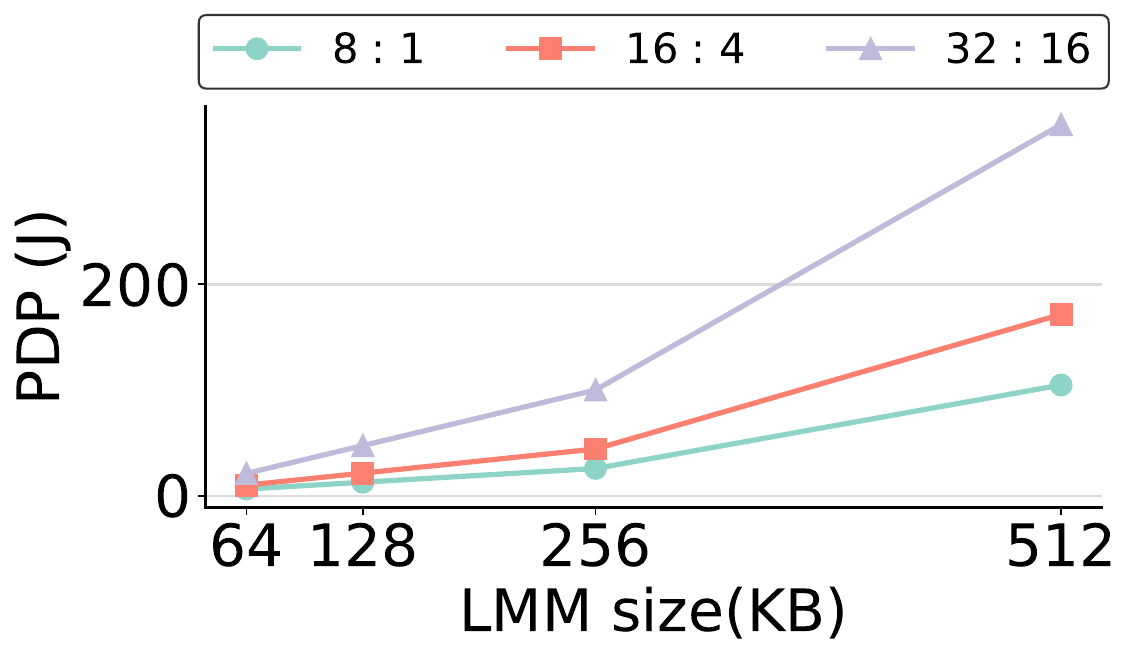}
  }\hspace{0.2em}
  \subfloat[Qwen3-0.6B Q8\_0\label{fig:sub_b}]{%
    \includegraphics[width=0.23\textwidth]{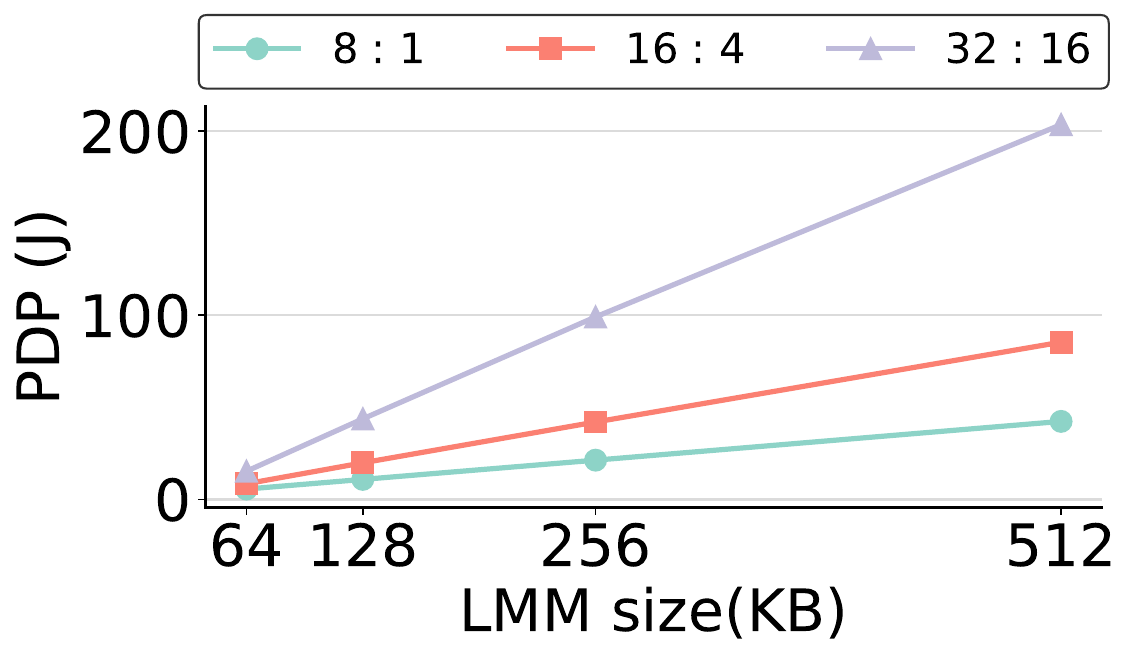}
  }\par\vspace{1em}
  \subfloat[Qwen3-1.7B Q3\_K\_S\label{fig:sub_c}]{%
    \includegraphics[width=0.23\textwidth]{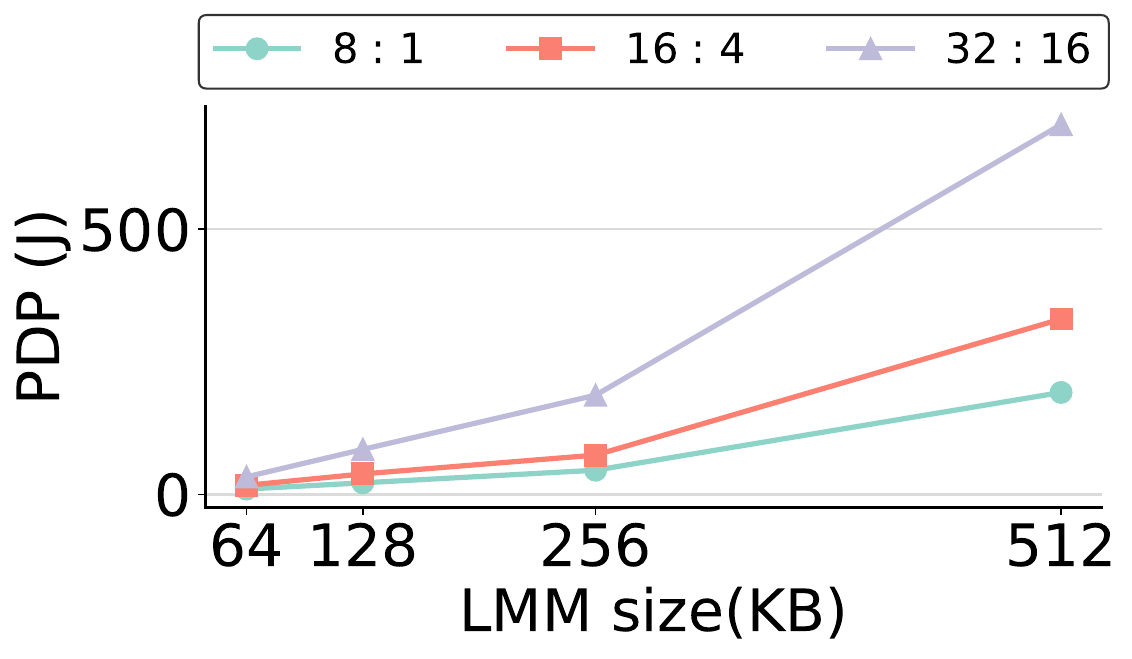}
  }\hspace{0.2em}
  \subfloat[Qwen3-1.7B Q8\_0\label{fig:sub_d}]{%
    \includegraphics[width=0.23\textwidth]{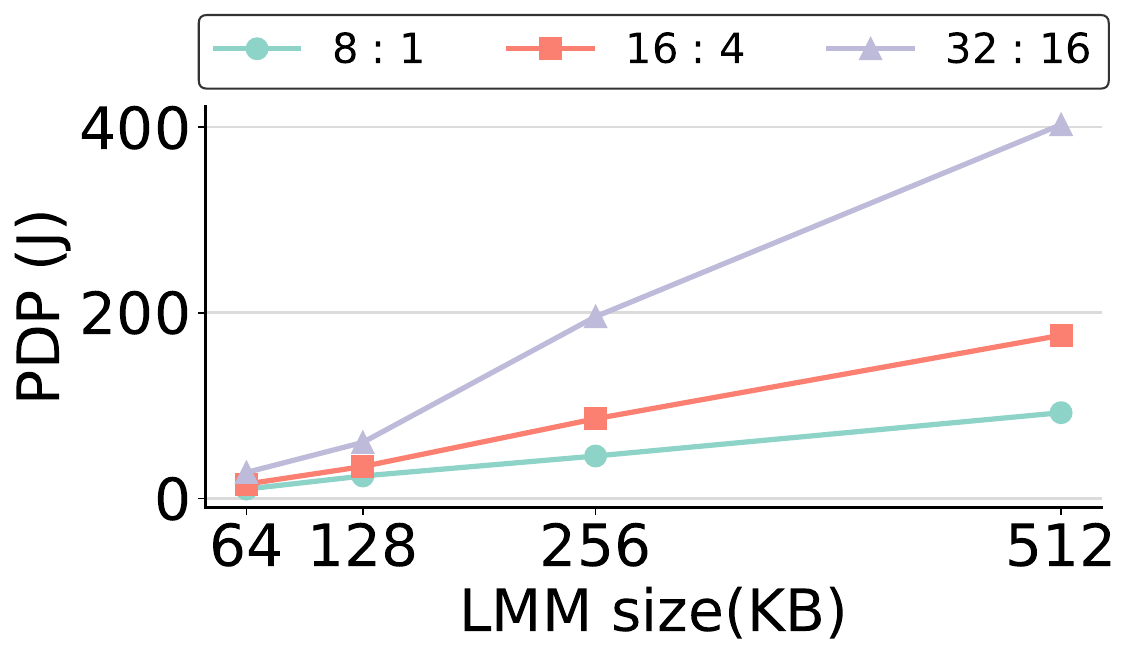}
  }\par\vspace{1em}
  \subfloat[Qwen3-8B Q3\_K\_S\label{fig:sub_e}]{%
    \includegraphics[width=0.23\textwidth]{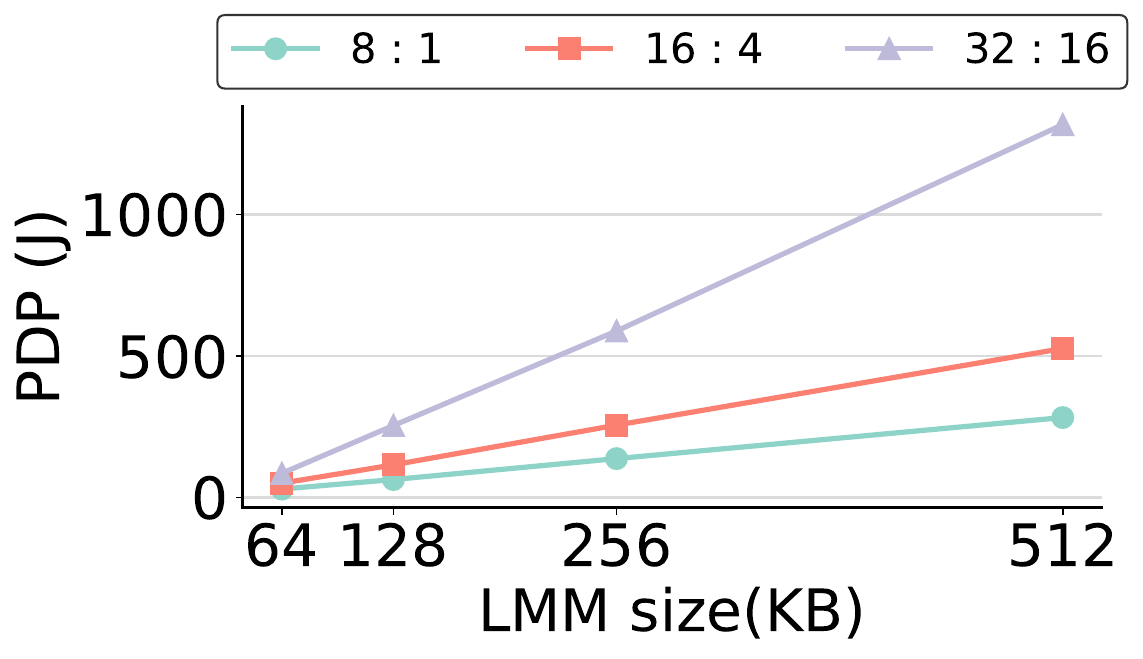}
  }\hspace{0.2em}
  \subfloat[Qwen3-8B Q8\_0\label{fig:sub_f}]{%
    \includegraphics[width=0.23\textwidth]{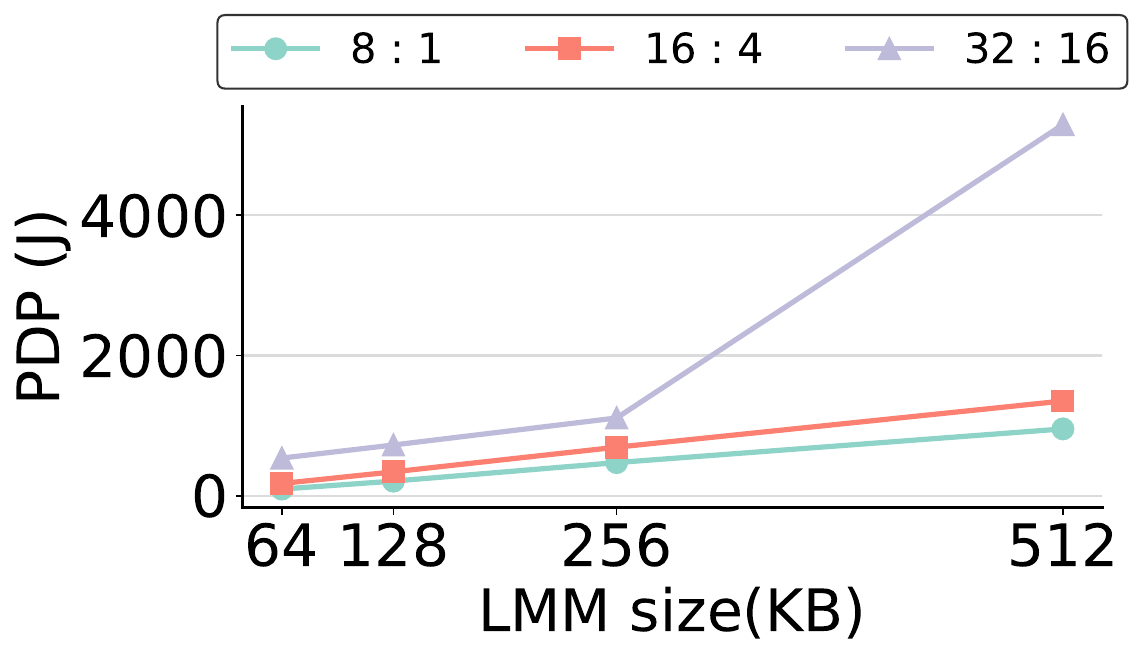}
  }
  \caption{Analysis of the impact of LMM size on energy efficiency (PDP, lower is better).}
  \label{fig:PDP_vs_LMM_matrix}
\end{figure}

In general, increasing the LMM size allows more computational kernels to reside entirely within on-chip memory. 
This strategy tends to improve the offload ratio, which is the proportion of operations executed on the IMAX accelerator rather than the host CPU, thereby reducing overall processing time. 
However, a larger LMM also linearly increases static power consumption, meaning that it does not automatically translate to better energy efficiency. 
Fig.~\ref{fig:PDP_vs_LMM_matrix} clearly illustrates this trade-off. 
For most workloads, increasing the LMM size beyond \SI{64}{K\byte} results in a higher PDP, as the penalty from increased power consumption outweighs the benefit of reduced execution time.
The offload ratios in Table~\ref{tab:model_scores_multirow} explain this phenomenon. 
For most models, a \SI{64}{K\byte} LMM is sufficient to achieve high offload ratios exceeding \SI{85}{\percent}, with the common FP16 kernel fitting entirely within this memory. 
This indicates that at \SI{64}{K\byte}, most targeted computations are already on-chip, offering limited potential for further improvement.
Consequently, the marginal performance improvement from larger LMMs fails to offset their increased power draw, leading to a degradation in PDP, making it more energy-efficient to execute it on the host.
The Qwen3-8B Q8\_0 model, while initially appearing to be an exception, reinforces this conclusion because the substantial size of its Q8\_0 kernel leads to excessive DMA transfer latency.
Consequently, offloading this kernel results in a higher PDP, as the overhead from data transfer outweighs the computational gains.
The most energy-efficient strategy, therefore, is to avoid offloading this specific kernel. 
This is precisely the behavior that a \SI{64}{K\byte} LMM enforces, making it an effective choice even for this challenging case.

\begin{table}[t]
  \centering
  \caption{Offload ratio of computational kernels to the IMAX for different Qwen3 models and quantization types.}
  \label{tab:model_scores_multirow}
  \begin{adjustbox}{max width=\columnwidth}
  \begin{tabular}{llrrrrr}
    \toprule
    Model & Quantized type & FP16 & Q3\_K & Q6\_K & Q8\_0 & Total \\
    \midrule
    \multirow{2}{*}{Qwen3-0.6B} & Q3\_K\_S & \SI{100}{\percent} & \SI{0}{\percent} & \SI{99.33}{\percent} & - & \SI{99.94}{\percent} \\
                                & Q8\_0  & \SI{100}{\percent} & - & - & \SI{89.44}{\percent} & \SI{91.13}{\percent} \\
    \midrule
    \multirow{2}{*}{Qwen3-1.7B} & Q3\_K\_S & \SI{100}{\percent} & \SI{99.91}{\percent} & \SI{0}{\percent} & - & \SI{94.27}{\percent} \\
                                & Q8\_0  & \SI{100}{\percent} & - & - & \SI{83.87}{\percent} & \SI{85.59}{\percent} \\
    \midrule
    \multirow{2}{*}{Qwen3-8B}   & Q3\_K\_S & \SI{100}{\percent} & \SI{89.09}{\percent} & \SI{0}{\percent} & - & \SI{88.23}{\percent} \\
                                & Q8\_0  & \SI{100}{\percent} & - & - & \SI{0}{\percent} & \SI{11.51}{\percent} \\
    \bottomrule
  \end{tabular}

  \end{adjustbox}

  \begin{minipage}{\columnwidth}
    \footnotesize
    \vspace{1em}
    \textbf{Note:} ``\SI{0}{\percent}'' indicates that offloading is possible but was not performed; ``-'' indicates that there is no computation to be offloaded for that kernel.
  \end{minipage}
\end{table}

\begin{figure}[t]
  \centering
  \subfloat[Prefill phase breakdown\label{fig:prefill_breakdown}]{%
    \includegraphics[width=\columnwidth]{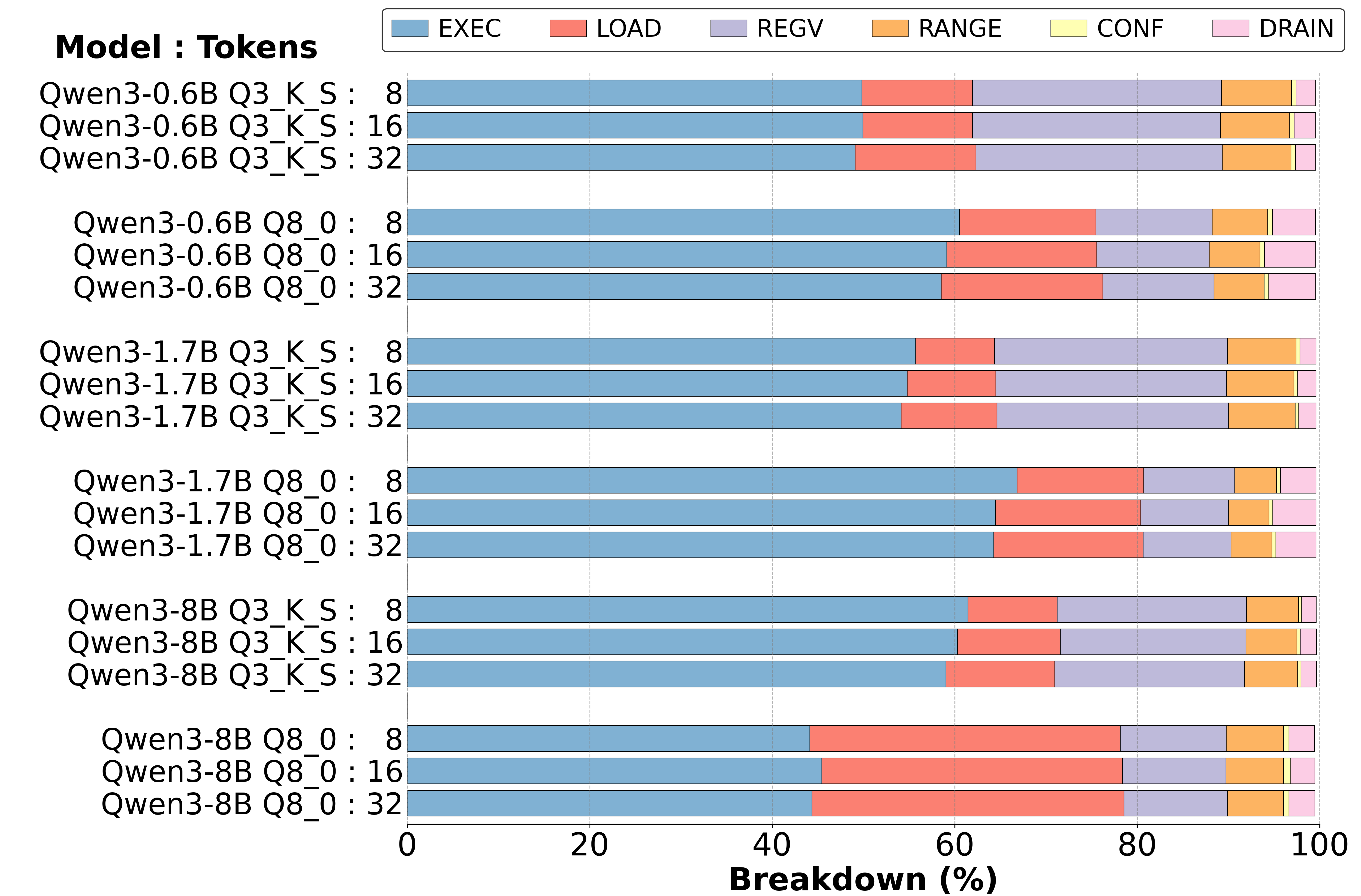}
  }\par\vspace{1em}
  \subfloat[Decode phase breakdown\label{fig:decode_breakdown}]{%
    \includegraphics[width=\columnwidth]{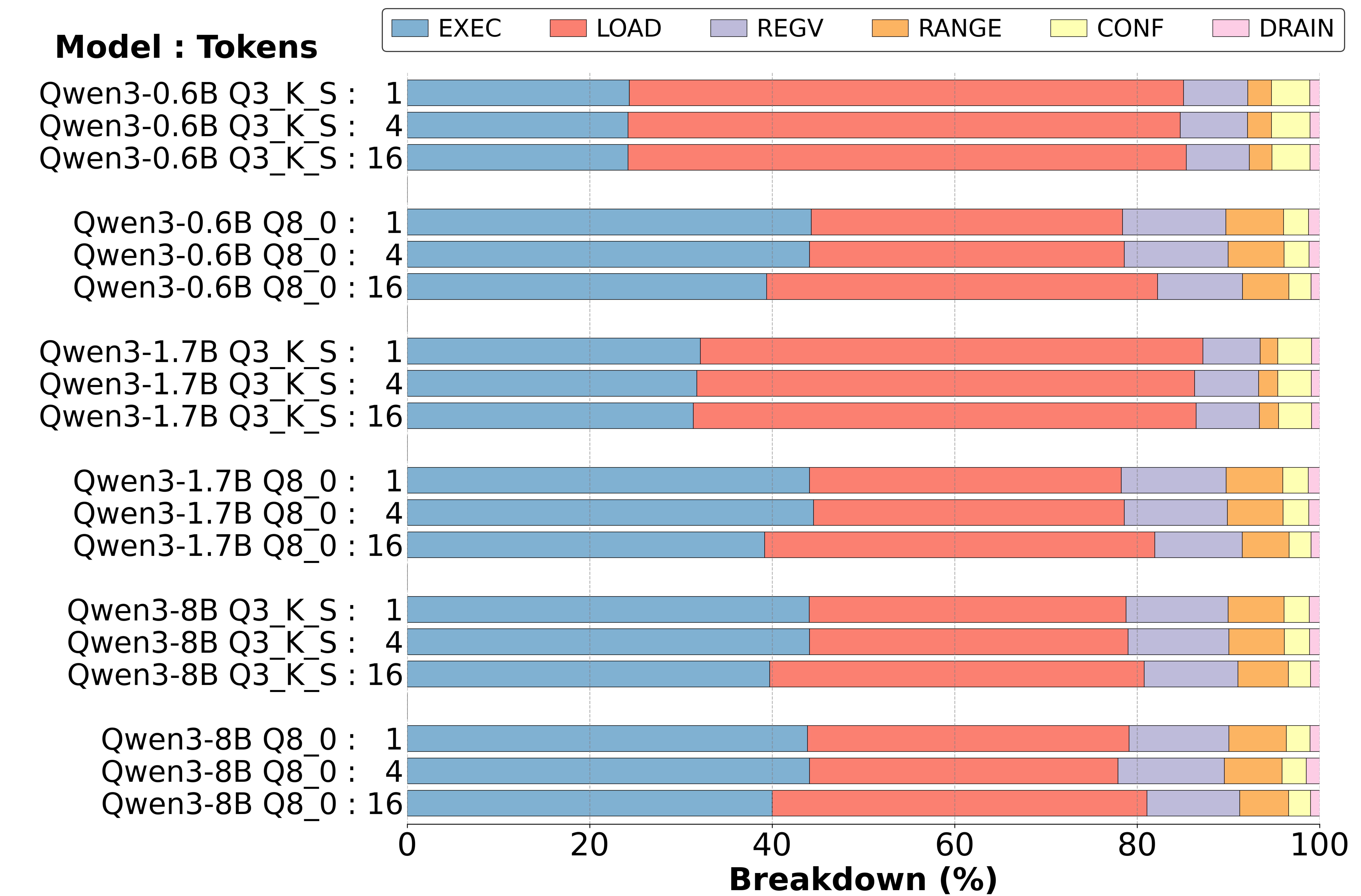}
  }
  \caption{Breakdown of execution time within the IMAX accelerator for the Prefill and Decode phases.}
  \label{fig:execution_time_breakdown}
\end{figure}

\subsection{Analysis of IMAX Execution Time Breakdown}

To identify performance bottlenecks, we conducted a multi-level breakdown of the IMAX execution time. 
We first analyze the E2E latency from a system-level (macro) perspective, and then analyze the internal (micro) execution phases within the IMAX accelerator.

A detailed breakdown of the E2E latency for a representative workload (Qwen3-0.6B Q3\_K\_S with a [32:16] token I/O) reveals that system-level overheads are the dominant performance bottleneck. 
Out of a total latency of 16.3 s, the actual IMAX kernel execution accounted for only \SI{4.47}{\second} (\SI{27.4}{\percent}). 
The majority of the time was consumed by host CPU processing and DMA data loading from host to IMAX, which required \SI{5.43}{\second} (\SI{33.3}{\percent}) and \SI{5.31}{\second} (\SI{32.6}{\percent}), respectively. 
The remaining time was spent on DMA data drain (\SI{0.31}{\second}, \SI{1.9}{\percent}) and other IMAX configuration tasks (\SI{0.78}{\second}, \SI{4.8}{\percent}). 
This quantitative analysis immediately highlights a critical finding: the time spent on DMA data loading (\SI{5.31}{\second}) is substantial, exceeding even the net kernel execution time (\SI{4.47}{\second}).
While a degree of host CPU overhead (\SI{33.3}{\percent}) is expected in a heterogeneous system for tasks such as scheduling and data preparation, the significant latency from the DMA load points to the host-accelerator data path constitutes a definitive system-level bottleneck.

This system-level bottleneck is a direct reflection of the accelerator's internal behavior. 
To understand this relationship, we now analyze the breakdown of the time spent within the offloaded tasks on IMAX. 
LLM inference is characterized by two distinct phases: the prefill phase, which processes the input prompt in parallel, and the decode phase, which generates tokens sequentially. 
As shown in Fig.~\ref{fig:execution_time_breakdown}, we categorize the execution time of each phase into six components:
\begin{itemize}
  \item \textbf{EXEC}: Kernel execution time on the IMAX cores.
  \item \textbf{LOAD}: Duration of input data transfer from host main memory to the LMMs via DMA.
  \item \textbf{DRAIN}: Duration of result data transfer from the LMMs back to host main memory via DMA.
  \item \textbf{CONF}:  Host CPU overhead for configuring mapping commands on IMAX via Programmed I/O (PIO).
  \item \textbf{REGV}: Host CPU overhead for initializing internal Processing Element (PE) registers on IMAX via PIO.
  \item \textbf{RANGE}: Host CPU overhead for configuring the LMM address space on IMAX via PIO.
\end{itemize}

The prefill phase is generally compute-bound.
For most workloads, excluding the Qwen3-8B Q8\_0 model, the net computation time (EXEC) accounts for over \SI{50}{\percent} of the total execution time. 
This indicates that the computational resources of IMAX are being utilized efficiently. 
However, we also observe a trend where the proportion of time spent on data transfer~(LOAD) increases with the number of input tokens, a direct consequence of the larger data sizes associated with longer prompts. 
This suggests that even in the prefill phase, data transfer can become a significant bottleneck for large-scale models.
Furthermore, we note that the proportion of time spent on result data transfer~(DRAIN) is larger in the decode phase compared to the prefill phase. 
This difference in DRAIN ratio reflects the distinct data characteristics of each phase, specifically KV cache generation in prefill versus single token output in decode. 
This observation confirms that the behavior of the CGLA architecture aligns with the inherent workload duality.
While our experiments cover typical interactive scenarios, the performance implications for workloads with much longer token sequences, such as large document summarization, also warrant consideration. 
The trends observed suggest that data transfer would become an even more significant bottleneck in such scenarios. 
In the prefill phase, a longer prompt linearly increases the LOAD duration, while in the decode phase, a longer context length requires loading an ever-larger KV cache. 
These factors can lead to memory bandwidth saturation, indicating that the increasing data transfer overhead is a critical bottleneck for scaling to very long contexts. 
This observation suggests that enhancing the host-accelerator interconnect will be a key consideration for improving scalability in future architectural designs.

In contrast, the decode phase exhibits a clear memory-bound characteristic and is specifically LOAD-bound. 
This behavior stems from the algorithmic nature of the decode phase. 
The computational workload is relatively small, as only a single token is generated per step. 
However, the entire large KV cache, generated from all previous tokens, must be loaded from memory in each iteration.
The LOAD time is the primary component of the overall execution. 
Its proportional share grows with longer context lengths because of the corresponding expansion of the KV cache.
Finally, a notable observation in the prefill breakdown for the Q3\_K\_S models is the significant proportion of register initialization overhead~(REGV). 
This is primarily attributed to the Q6\_K kernel, which utilizes all 64 PEs of the IMAX architecture and thus requires a substantial amount of register configuration. 
The prefill phase's higher utilization of the Q6\_K kernel results in this increased REGV overhead.

These measurements were taken with IMAX's double-buffered LMMs actively overlapping computation and DMA transfers. 
That data transfer remains the dominant bottleneck, even with this hardware optimization, highlights the severity of the memory bandwidth challenge in LLM inference. 
While architectural overlap is essential, it is insufficient, indicating that further optimizations at both the system and algorithmic levels are required.

\begin{figure}[t]
  \centering
  \includegraphics[width=1\columnwidth]{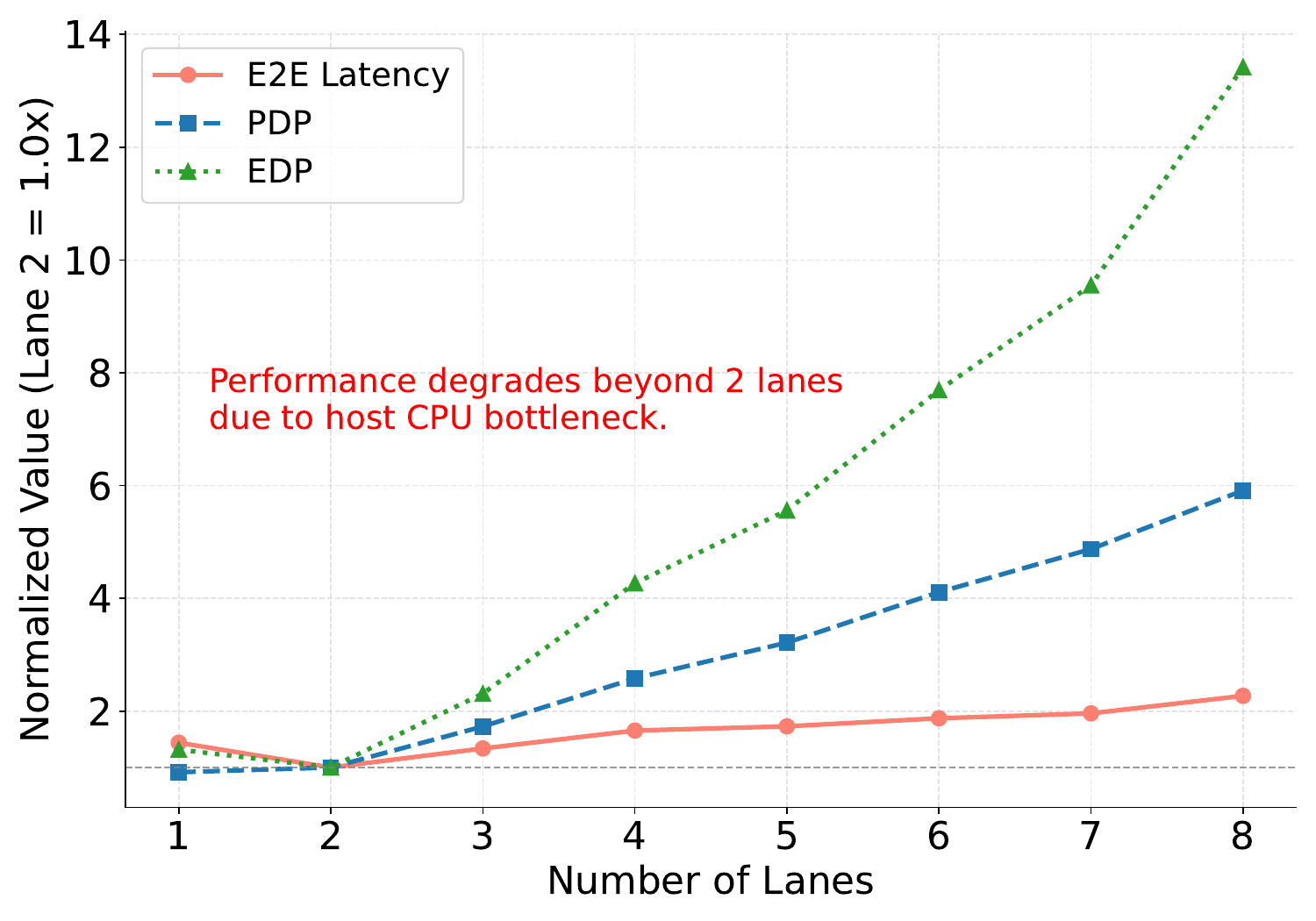}
  \caption{Scalability analysis of IMAX performance metrics as the number of compute lanes increases.}
  \label{fig:performance_metrics_normalized}

\end{figure}

\subsection{System-Level Limitations and Future Directions}
This work demonstrates the potential of a general-purpose CGRA for energy-efficient LLM inference, but it is equally important to discuss the limitations of our approach, which in turn define critical directions for future work.

A primary limitation, as revealed by our scalability analysis, is the architecture's dependence on the host system's performance.
As shown in Fig.~\ref{fig:performance_metrics_normalized}, performance saturates and then degrades beyond a two-lane configuration. 
This bottleneck is not inherent to the IMAX architecture itself but is a direct consequence of the dual-core ARM host's limited capability to manage data transfers and control flow for multiple parallel lanes. 
This finding highlights a fundamental challenge in heterogeneous computing, where an accelerator's performance is often constrained by its host system.
Therefore, a crucial next step is to integrate the IMAX architecture with a higher-performance host (e.g., an eight-core CPU), ideally via a high-bandwidth PCIe interconnect, to experimentally validate its true scalability potential.

Further limitations relate to the scope of our evaluation.
First, the performance and power figures for the \SI{28}{\nano\meter} ASIC are projections derived from synthesis tools. 
While based on standard industry practice, these estimates are subject to variations in the final physical implementation and manufacturing process. 
Second, our experiments were conducted on a specific embedded platform (AMD Versal VPK180 FPGA). 
System dynamics, particularly data transfer overhead, may differ significantly in other environments, such as a server with a PCIe-based interconnect. 
Finally, our analysis focused on the Qwen3 model family. 
Although our approach theoretically supports larger model sizes, the prototype's limited DMA buffer size restricted our experiments to the current configurations.
While representative, future work should extend this evaluation to models with diverse architectures, such as MoE, to ensure broader applicability.

Addressing these limitations forms a clear approach for future research. 
Beyond scaling the host system, optimizing the host-accelerator interface through software co-design and exploring hardware support for emerging low-bit quantization formats remain promising approaches to further enhance performance and efficiency.

%% file: conclusion.tex
\section{Conclusion}
\label{conclusions}

In this paper, we investigated the effectiveness of the general-purpose CGLA accelerator, IMAX, in addressing the fundamental challenge of high energy consumption in LLM inference. 
We presented the first implementation of the state-of-the-art Qwen3 LLM family, along with a diverse set of quantized kernels, on IMAX using the practical llama.cpp framework. 
We then conducted a comprehensive E2E performance evaluation based on an FPGA prototype and a projected \SI{28}{\nano\meter} ASIC implementation.  
Although the proposed approach does not match the E2E latency of GPGPUs, the experimental results demonstrate its superior energy efficiency. 
The projected IMAX ASIC achieves up to a \num{44.4}$\times$ improvement in PDP and a \num{11.5}$\times$ improvement in EDP compared to the NVIDIA RTX 4090.
These results demonstrate that a general-purpose architecture can attain high energy efficiency on modern LLM inference tasks.

The analysis of the primary bottlenecks identified in this work provides clear architectural guidance for future designs. 
Our findings highlight the critical need to redesign the host-accelerator interface for data-center-scale performance and to explore co-design opportunities with emerging low-bit quantization formats. 
These promising approaches represent key directions for future research.

%% file: bibliography.bib
@IEEEtranBSTCTL{IEEEtranBSTCTL,
  CTLuse_forced_etal       = "yes",
  CTLname_show_etal        = "1",
  CTLnames_truncate_to     = "5",
  CTLuse_url = "no"
}

@inproceedings{fpgacnn,
  author    = {C. Zhang and P. Li and G. Sun and Y. Guan and B. Xiao and J. Cong},
  title     = {{O}ptimizing {FPGA}-based accelerator design for deep convolutional neural networks},
  booktitle = {Proc. ACM/SIGDA Int. Symp. Field-Programmable Gate Arrays (FPGA)},
  year      = {2015},
  pages     = {161--170},
  doi       = {10.1145/2684746.2689060}
}

@misc{llama.cpp,
  author       = {G. Gerganov},
  title        = {{Llama.cpp}: {LLM} inference in {C/C++}},
  year         = {2023},
  organization = {GitHub},
  howpublished = {[Online]. Available: \url{https://github.com/ggerganov/llama.cpp}},
  note         = {Accessed: May 11, 2025} 
}

@misc{SynopsysNDDCUltra,
  author       = {{Synopsys, Inc.}},
  organization = {{Synopsys}},
  title        = "{DC Ultra}: Concurrent timing, area, power and test optimization",
  howpublished = {[Online]. Available: \url{https://www.synopsys.com/implementation-and-signoff/rtl-synthesis-test/dc-ultra.html}},
  note         = {Accessed: May 25, 2025}
}

@misc{li2020ftransenergyefficientaccelerationtransformers,
      title={{FTRANS}: Energy-Efficient Acceleration of Transformers using FPGA}, 
      author={Bingbing Li and Santosh Pandey and Haowen Fang and Yanjun Lyu and Ji Li and Jieyang Chen and Mimi Xie and Lipeng Wan and Hang Liu and Caiwen Ding},
      year={2020},
      note = {arXiv:2007.08563}
}

@inproceedings{vitisai,
  author={J. Wang and S. Gu},
  booktitle={Proc. 11th Int. Conf. Inf. Sci. Technol. (ICIST)},
  title="{FPGA} implementation of object detection accelerator based on {Vitis-AI}",
  year={2021},
  pages={571-577},
  doi={10.1109/ICIST52614.2021.9440554}}

@INPROCEEDINGS{multi_task,
  author={Lu, Yufan and Zhai, Xiaojun and Saha, Sangeet and Ehsan, Shoaib and McDonald-Maier, Klaus D.},
  booktitle={Proc. 14th IEEE Int. Symp. Embedded Multicore/Many-Core Syst.-on-Chip (MCSoC)},
  title={{FPGA} based Adaptive Hardware Acceleration for Multiple Deep Learning Tasks}, 
  year={2021},
  pages={204-209},
  doi={10.1109/MCSoC51149.2021.00038}}

@misc{Intel_Xeon_w5-2455X_Specs,
  author = {{Intel}},
  organization = {{Intel}},
  title = "{{Intel{\textregistered}} {Xeon{\textregistered}} w5-2455{X} processor (30{M} cache, 3.20 {G}{Hz}) specifications}",
  howpublished = {[Online]. Available: \url{https://www.intel.co.jp/content/www/jp/ja/products/sku/233420/intel-xeon-w52455x-processor-30m-cache-3-20-ghz/specifications.html}},
  note  = {Accessed: May 30, 2025}
}

@misc{iea_energy_ai,
  author       = {{International Energy Agency}},
  title        = "{E}nergy and {AI}",
  year         = {2025},
   organization = {IEA},
  howpublished = {[Online]. Available: \url{https://www.iea.org/reports/energy-and-ai}},
  note         = {Accessed: Jun. 14, 2025}
}

@misc{Versal,
  author       = {{Xilinx}},
  title        = {Versal power demo},
  organization = {{Xilinx}},
  howpublished = {[Online]. Available: \url{https://github.com/Xilinx/pm_demo/tree/master?tab=readme-ov-file}},
  note         = {Accessed: May 30, 2025}
}

@techreport{nvidia_1080ti,
  author = {Nvidia},
  title = {{NVIDIA} Turing {GPU} Architecture},
  institution = {NVIDIA},
  address     = {Santa Clara, CA, USA},
  type = {White Paper},
  year = {2018}
}

@techreport{nvidia_4090,
  author = {{Nvidia Corporation}},
  title = "{NVIDIA} {ADA} {GPU} architecture: {D}esigned to deliver outstanding gaming and creating, professional graphics, {AI}, and compute performance",
  institution = {{Nvidia Corporation}},
  type = {White Paper},
  address     = {Santa Clara, CA, USA}, 
  year = {2023}
}

@techreport{nvidia_jetson_agx_orin,
  title="{Nvidia} {J}etson {AGX} {O}rin series: A giant leap forward for robotics and edge {AI} applications",
  author={L. S. Karumbunathan},
  institution={NVIDIA},
  type={Tech. Brief},
  address     = {Santa Clara, CA, USA}, 
  year={2022}
}

@misc{LLM_survey1,
      title={Large Language Models: A Survey},
      author  = {Minaee, Shervin and Mikolov, Tomas and Nikzad, Narjes and Chenaghlu, Meysam and Socher, Richard and Amatriain, Xavier and Gao, Jianfeng},
      year={2024},
      eprint={2402.06196},
      note={arXiv:2402.06196},
      archivePrefix={arXiv},
      primaryClass={cs.CL}
}

@misc{LLM_survey2,
      title={A Survey on {LLM}-as-a-Judge}, 
      author={Jiawei Gu and Xuhui Jiang and Zhichao Shi and Hexiang Tan and Xuehao Zhai and Chengjin Xu and Wei Li and Yinghan Shen and Shengjie Ma and Honghao Liu and Saizhuo Wang and Kun Zhang and Yuanzhuo Wang and Wen Gao and Lionel Ni and Jian Guo},
      year={2025},
      eprint={2411.15594},
      archivePrefix={arXiv},
      primaryClass={cs.CL},
      note={arXiv:2411.15594}
}

@ARTICLE{LLM_survey3,
  author={Siino, Marco and Falco, Mariana and Croce, Daniele and Rosso, Paolo},
  journal={IEEE Access}, 
  title={Exploring {LLMs} Applications in Law: A Literature Review on Current Legal {NLP} Approaches}, 
  year={2025},
  volume={13},
  number={},
  pages={18253-18276},
  doi={10.1109/ACCESS.2025.3533217}}

@INPROCEEDINGS{LLM_survey4,
  author={Sindhu,B  and Prathamesh,R P and Sameera ,M B and KumaraSwamy,S},
  booktitle={Proc. Int. Conf. Current Trends Adv. Comput. (ICCTAC)},
  title={The Evolution of Large Language Model: Models, Applications and Challenges}, 
  year={2024},
  pages={1-8},
  doi={10.1109/ICCTAC61556.2024.10581180}}

@article{LLM_survey5,
    author = {Wu, Junchao and Yang, Shu and Zhan, Runzhe and Yuan, Yulin and Chao, Lidia Sam and Wong, Derek Fai},
    title = {A Survey on {LLM}-Generated Text Detection: Necessity, Methods, and Future Directions},
    journal = {Computational Linguistics},
    volume = {51},
    number = {1},
    pages = {275--338},
    year = {2025},
    month = mar,
    issn = {0891-2017},
    doi = {10.1162/coli\_a\_00549}
}

@article{NLP_survey1, 
        title={A Comparative Analysis of Generative Artificial Intelligence Tools for Natural Language Processing}, 
        volume={1},
        doi={10.62411/jcta.9447},
        number={3},
        journal={J. Comput. Theor. Appl.},
        author={A. Iorliam and J. A. Ingio},
        year={2024},
        month={Feb.},
        pages={311-325}
}

@inproceedings{NLP_survey2,
author = {Karanikolas, Nikitas and Manga, Eirini and Samaridi, Nikoletta and Tousidou, Eleni and Vassilakopoulos, Michael},
title = {Large language models versus natural language understanding and generation},
year = {2024},
booktitle = {Proc. 27th Pan-Hellenic Conf. Progress Comput. Inform.},
pages = {278-290},
doi = {10.1145/3635059.3635104}
}

@ARTICLE{NLP_survey3,
  author={Nasution, Arbi Haza and Onan, Aytuğ},
  journal={IEEE Access},
  title="{ChatGPT} label: Comparing the quality of human-generated and {LLM}-generated annotations in low-resource language {NLP} tasks",
  year={2024},
  volume={12},
  pages={71876-71900},
  doi={10.1109/ACCESS.2024.3402809}}

@INPROCEEDINGS{code_generation1,
  author={Wang, Jianxun and Chen, Yixiang},
  booktitle={Proc. IEEE Int. Conf. Med. Artif. Intell. (MedAI)},
  title={A review on code generation with {LLMs}: Application and evaluation},
  year={2023},
  pages={284-289},
  doi={10.1109/MedAI59581.2023.00044}}

@article{code_generation2,
author = {Huang, Dong and Zhang, Jie M. and Bu, Qingwen and Xie, Xiaofei and Chen, Junjie and Cui, Heming},
title = {Bias testing and mitigation in {LLM}-based code generation},
year = {2025},
journal = {ACM Trans. Softw. Eng. Methodol.},
doi = {10.1145/3724117}
}

@inproceedings{code_generation3,
author = {Wermelinger, Michel},
title = "{U}sing {GitHub} {C}opilot to solve simple programming problems",
year = {2023},
booktitle = {Proc. 54th ACM Tech. Symp. Comput. Sci. Educ.},
pages = {172-178},
doi = {10.1145/3545945.3569830}
}

@ARTICLE{conversational_agents1,
  author={Kusal, Sheetal and Patil, Shruti and Choudrie, Jyoti and Kotecha, Ketan and Mishra, Sashikala and Abraham, Ajith},
  journal={IEEE Access}, 
  title={{AI-based conversational agents: A scoping review from technologies to future directions}}, 
  year={2022},
  volume={10},
  pages={92337-92356},
  doi={10.1109/ACCESS.2022.3201144}}

@article{conversational_agents2,
author = {Mo, Fengran and Mao, Kelong and Zhao, Ziliang and Qian, Hongjin and Chen, Haonan and Cheng, Yiruo and Li, Xiaoxi and Zhu, Yutao and Dou, Zhicheng and Nie, Jian-Yun},
title = "{A} survey of conversational search",
  journal = {{ACM} Trans. Inf. Syst.},
  year    = {2025},
  month   = aug,
  doi     = {10.1145/3759453},
  note    = {early access}
}

@inproceedings{multi-modal1, 
    title={{BLIVA}: A Simple Multimodal {LLM} for Better Handling of Text-Rich Visual Questions}, 
    volume={38}, 
    DOI={10.1609/aaai.v38i3.27999}, 
    number={3}, 
    booktitle={Proc. AAAI Conf. Artif. Intell.},
    author={Hu, Wenbo and Xu, Yifan and Li, Yi and Li, Weiyue and Chen, Zeyuan and Tu, Zhuowen}, 
    year={2024}, 
    month={Mar.}, 
    pages={2256-2264} 
}

@INPROCEEDINGS {multi-modal4,
author = { Cui, Can and Ma, Yunsheng and Cao, Xu and Ye, Wenqian and Zhou, Yang and Liang, Kaizhao and Chen, Jintai and Lu, Juanwu and Yang, Zichong and Liao, Kuei-Da and Gao, Tianren and Li, Erlong and Tang, Kun and Cao, Zhipeng and Zhou, Tong and Liu, Ao and Yan, Xinrui and Mei, Shuqi and Cao, Jianguo and Wang, Ziran and Zheng, Chao },
booktitle = {Proc. IEEE/CVF Winter Conf. Appl. Comput. Vis. Workshops (WACVW)},
title = "{A} survey on multimodal large language models for autonomous driving",
year = {2024},
pages = {958-979},
doi = {10.1109/WACVW60836.2024.00106},
publisher = {IEEE Comput. Soc.},
address = {Los Alamitos, CA, USA},
month = jan}

@misc{multi-modal5,
      title={{Macaw-LLM}: Multi-Modal Language Modeling with Image, Audio, Video, and Text Integration}, 
      author={Chenyang Lyu and Minghao Wu and Longyue Wang and Xinting Huang and Bingshuai Liu and Zefeng Du and Shuming Shi and Zhaopeng Tu},
      year={2023},
      eprint={2306.09093},
      note={arXiv:2306.09093}
}

@article{gpu_power1,
author = {Bridges, Robert A. and Imam, Neena and Mintz, Tiffany M.},
title = {Understanding {GPU} power: A survey of profiling, modeling, and simulation methods},
year = {2016},
publisher = {Assoc. Comput. Mach.},
address = {New York, NY, USA},
volume = {49},
number = {3},
doi = {10.1145/2962131},
journal = {ACM Comput. Surv.},
month = sep
}

@article{gpu_power2,
   title={Trends in {AI} inference energy consumption: Beyond the performance-vs-parameter laws of deep learning},
   volume={38},
   doi={10.1016/j.suscom.2023.100857},
   journal={Sustain. Comput., Inform. Syst.},
   author={R. Desislavov and F. Martínez-Plumed and J. Hernández-Orallo},
   year={2023},
   month=apr,
   pages   = {100857}
}

@misc{yang2025qwen3technicalreport,
      title={Qwen3 technical report}, 
      author={An Yang and Anfeng Li and Baosong Yang and Beichen Zhang and Binyuan Hui and Bo Zheng and Bowen Yu and Chang Gao and Chengen Huang and Chenxu Lv and Chujie Zheng and Dayiheng Liu and Fan Zhou and Fei Huang and Feng Hu and Hao Ge and Haoran Wei and Huan Lin and Jialong Tang and Jian Yang and Jianhong Tu and Jianwei Zhang and Jianxin Yang and Jiaxi Yang and Jing Zhou and Jingren Zhou and Junyang Lin and Kai Dang and Keqin Bao and Kexin Yang and Le Yu and Lianghao Deng and Mei Li and Mingfeng Xue and Mingze Li and Pei Zhang and Peng Wang and Qin Zhu and Rui Men and Ruize Gao and Shixuan Liu and Shuang Luo and Tianhao Li and Tianyi Tang and Wenbiao Yin and Xingzhang Ren and Xinyu Wang and Xinyu Zhang and Xuancheng Ren and Yang Fan and Yang Su and Yichang Zhang and Yinger Zhang and Yu Wan and Yuqiong Liu and Zekun Wang and Zeyu Cui and Zhenru Zhang and Zhipeng Zhou and Zihan Qiu},
      year={2025},
      eprint={2505.09388},
      archivePrefix={arXiv},
      primaryClass={cs.CL},
      note={arXiv:2505.09388}
}

@misc{qwen,
      title={Qwen Technical Report}, 
      author={Jinze Bai and Shuai Bai and Yunfei Chu and Zeyu Cui and Kai Dang and Xiaodong Deng and Yang Fan and Wenbin Ge and Yu Han and Fei Huang and Binyuan Hui and Luo Ji and Mei Li and Junyang Lin and Runji Lin and Dayiheng Liu and Gao Liu and Chengqiang Lu and Keming Lu and Jianxin Ma and Rui Men and Xingzhang Ren and Xuancheng Ren and Chuanqi Tan and Sinan Tan and Jianhong Tu and Peng Wang and Shijie Wang and Wei Wang and Shengguang Wu and Benfeng Xu and Jin Xu and An Yang and Hao Yang and Jian Yang and Shusheng Yang and Yang Yao and Bowen Yu and Hongyi Yuan and Zheng Yuan and Jianwei Zhang and Xingxuan Zhang and Yichang Zhang and Zhenru Zhang and Chang Zhou and Jingren Zhou and Xiaohuan Zhou and Tianhang Zhu},
      year={2023},
      eprint={2309.16609},
      archivePrefix={arXiv},
      primaryClass={cs.CL},
      note={arXiv:2309.16609}
}

@misc{TPU,
      title={{TPU v4}: An Optically Reconfigurable Supercomputer for Machine Learning with Hardware Support for Embeddings}, 
      author={Norman P. Jouppi and George Kurian and Sheng Li and Peter Ma and Rahul Nagarajan and Lifeng Nai and Nishant Patil and Suvinay Subramanian and Andy Swing and Brian Towles and Cliff Young and Xiang Zhou and Zongwei Zhou and David Patterson},
      year={2023},
      eprint={2304.01433},
      note={arXiv:2304.01433}
}

@inproceedings{MTIA,
author = {Firoozshahian, Amin and Coburn, Joel and Levenstein, Roman and Nattoji, Rakesh and Kamath, Ashwin and Wu, Olivia and Grewal, Gurdeepak and Aepala, Harish and Jakka, Bhasker and Dreyer, Bob and Hutchin, Adam and Diril, Utku and Nair, Krishnakumar and Aredestani, Ehsan K. and Schatz, Martin and Hao, Yuchen and Komuravelli, Rakesh and Ho, Kunming and Abu Asal, Sameer and Shajrawi, Joe and Quinn, Kevin and Sreedhara, Nagesh and Kansal, Pankaj and Wei, Willie and Jayaraman, Dheepak and Cheng, Linda and Chopda, Pritam and Wang, Eric and Bikumandla, Ajay and Karthik Sengottuvel, Arun and Thottempudi, Krishna and Narasimha, Ashwin and Dodds, Brian and Gao, Cao and Zhang, Jiyuan and Al-Sanabani, Mohammed and Zehtabioskuie, Ana and Fix, Jordan and Yu, Hangchen and Li, Richard and Gondkar, Kaustubh and Montgomery, Jack and Tsai, Mike and Dwarakapuram, Saritha and Desai, Sanjay and Avidan, Nili and Ramani, Poorvaja and Narayanan, Karthik and Mathews, Ajit and Gopal, Sethu and Naumov, Maxim and Rao, Vijay and Noru, Krishna and Reddy, Harikrishna and Venkatapuram, Prahlad and Bjorlin, Alexis},
title = {{MTIA}: First Generation Silicon Targeting {Meta's} Recommendation Systems},
year = {2023},
isbn = {9798400700958},
publisher = {Association for Computing Machinery},
address = {New York, NY, USA},
doi = {10.1145/3579371.3589348},
booktitle = {Proceedings of the 50th Annual International Symposium on Computer Architecture},
articleno = {80},
numpages = {13},
location = {Orlando, FL, USA},
series = {ISCA '23}
}

@inproceedings{MoE,
 author = {Riquelme, Carlos and Puigcerver, Joan and Mustafa, Basil and Neumann, Maxim and Jenatton, Rodolphe and Susano Pinto, Andr\'{e} and Keysers, Daniel and Houlsby, Neil},
 booktitle = {Advances in Neural Information Processing Systems},
 editor = {M. Ranzato and A. Beygelzimer and Y. Dauphin and P.S. Liang and J. Wortman Vaughan},
 pages = {8583--8595},
 publisher = {Curran Associates, Inc.},
 title = {Scaling Vision with Sparse Mixture of Experts},
 volume = {34},
 year = {2021}
}

@misc{AWQ,
      title={AWQ: Activation-aware Weight Quantization for LLM Compression and Acceleration}, 
      author={Ji Lin and Jiaming Tang and Haotian Tang and Shang Yang and Wei-Ming Chen and Wei-Chen Wang and Guangxuan Xiao and Xingyu Dang and Chuang Gan and Song Han},
      year={2024},
      archivePrefix={arXiv},
      primaryClass={cs.CL},
      note={arXiv:2306.00978}
}

@ARTICLE{imax3,
  author={Akabe, Tomoya and Trung Duong LE, Vu and Nakashima, Yasuhiko},
  journal={IEEE Access}, 
  title={{IMAX}: A Power-Efficient Multilevel Pipelined {CGLA} and Applications}, 
  year={2025},
  volume={13},
  number={},
  pages={31899-31911},
  doi={10.1109/ACCESS.2024.3524415}}

@INPROCEEDINGS {first_llm_imax,
author = { Uetani, Hitoaki and Nakashima, Yasuhiko },
booktitle = { Proc. 12th Int. Symp. Comput. Netw. (CANDAR) },
title = "{I}mplementation and evaluation of {LLM} on a {CGLA}",
year = {2024},
pages = {252-258},
doi = {10.1109/CANDAR64496.2024.00040}
}

@misc{llama2_imax,
  title={Implementation and performance analysis of {LLaMA} on a {CGLA}},
  author={Eto, Yu and Nakashima, Yasuhiko},
  note={Int. Conf. Intell. Syst. Netw. (ICISN), 2025}
}

@INPROCEEDINGS{cnn_imax_1,
  author={Tanomoto, Masakazu and Takamaeda, Shinya and Yao, Jun and Nakashima, Yasuhiko},
    booktitle={Proc. 9th IEEE Int. Symp. Embedded Multicore/Many-Core Syst.-on-Chip},
  title={A {CGRA}-based approach for accelerating convolutional neural networks},
  year={2015},
  pages={73-80},
  doi={10.1109/MCSoC.2015.41}}

@ARTICLE{cnn_imax_2,
  author={Thi Sang, Duong and Imamura, Ren and Akabe, Tomoya and Nakashima, Yasuhiko},
  journal={IEEE Access}, 
  title={Energy consumption optimization of multi-dimensional {U}-Nets on {CGLA}}, 
  year={2025},
  volume={13},
  number={},
  pages={29476-29492},
  doi={10.1109/ACCESS.2025.3539417}}

@INPROCEEDINGS{gat_imax,
  author={Asahina, Koki and Kim, Dohyun and Akabe, Tomoya and Duong Le, Vu Trung and Nakashima, Yasuhiko},
  booktitle={2025 International Conference on Machine Learning and Autonomous Systems (ICMLAS)}, 
  booktitle={Proc. Int. Conf. Mach. Learn. Auton. Syst. (ICMLAS)},
  title={Energy-efficient {SpMM} kernels for {GATs} and {GCNs} on a {CGLA}},
  year={2025},
  pages={1722-1729},
  doi={10.1109/ICMLAS64557.2025.10968638}}

@ARTICLE{knn_imax,
  author={Kim, Dohyun and Nakashima, Yasuhiko},
  journal={IEEE Access}, 
  title={Optimizing matrix-vector operations with {CGLA} for high-performance approximate k-NN search}, 
  year={2025},
  volume={13},
  number={},
  pages={111087-111097},
  doi={10.1109/ACCESS.2025.3582825}}

@INPROCEEDINGS{light_field_imax,
  author={Yuttakonkit, Yuttakon and Nakashima, Yasuhiko},
  booktitle={2016 Fourth Int. Symposium on Computing and Networking (CANDAR)}, 
  title={Performance Comparison of {CGRA} and Mobile {GPU} for Light-Field Image Processing}, 
  year={2016},
  volume={},
  number={},
  pages={174-180},
  doi={10.1109/CANDAR.2016.0040}}

@INPROCEEDINGS {bitmod,
author = { Chen, Yuzong and AbouElhamayed, Ahmed F. and Dai, Xilai and Wang, Yang and Andronic, Marta and Constantinides, George A. and Abdelfattah, Mohamed S. },
booktitle = { Proc. IEEE Int. Symp. High Perform. Comput. Archit. (HPCA) },
title = "{BitMoD}: Bit-serial mixture-of-datatype {LLM} acceleration",
year = {2025},
pages = {1082-1097},
doi = {10.1109/HPCA61900.2025.00084},
month =mar}

@misc{acc_llm,
      title={{AccLLM}: Accelerating Long-Context {LLM} Inference Via Algorithm-Hardware Co-Design}, 
      author={Yanbiao Liang and Huihong Shi and Haikuo Shao and Zhongfeng Wang},
      year={2025},
      eprint={2505.03745},
      note={arXiv:2505.03745}
}

@misc{flex_cim,
      title={Accelerating {LLM} Inference with Flexible N:M Sparsity via A Fully Digital Compute-in-Memory Accelerator}, 
      author={Akshat Ramachandran and Souvik Kundu and Arnab Raha and Shamik Kundu and Deepak K. Mathaikutty and Tushar Krishna},
      year={2025},
      eprint={2504.14365},
      note={arXiv:2504.14365}
}

@misc{secda-llm,
      title={Designing efficient {LLM} accelerators for edge devices}, 
      author={Jude Haris and Rappy Saha and Wenhao Hu and José Cano},
      year={2024},
      eprint={2408.00462},
      note={arXiv:2408.00462}
}

@misc{qwen3_quantization,
      title={An Empirical Study of Qwen3 Quantization}, 
      author={Xingyu Zheng and Yuye Li and Haoran Chu and Yue Feng and Xudong Ma and Jie Luo and Jinyang Guo and Haotong Qin and Michele Magno and Xianglong Liu},
      year={2025},
      eprint={2505.02214},
      archivePrefix={arXiv},
      primaryClass={cs.LG},
      note={arXiv:2505.02214}
}

@INPROCEEDINGS {ultraformer,
author = { Agostinelli, Victor and Agostini, Nicolas Bohm and Tumeo, Antonino },
booktitle = { Proc. 33rd Annu. IEEE Int. Symp. Field-Programmable Custom Comput. Mach. (FCCM) },
title = "{UltraFormer}: An efficient transformer for {FPGAs}",
year = {2025},
pages = {274-274},
doi = {10.1109/FCCM62733.2025.00034},
month =may}

@INPROCEEDINGS{hydra,
  author={Yang, Yinghao and Xu, Xicheng and Zhang, Haibin and Song, Jie and Tang, Xin and Lu, Hang and Li, Xiaowei},
  booktitle={Proc. IEEE Int. Symp. High Perform. Comput. Archit. (HPCA)},
  title={{Hydra}: Scale-out {FHE} accelerator architecture for secure deep learning on {FPGA}},
  year={2025},
  pages={1174-1186},
  doi={10.1109/HPCA61900.2025.00090}}

@INPROCEEDINGS{llm_agile,
  author = {Chen, Lvcheng and Wu, Ying and Wen, Chenyi and Wang, Shizhang and Zhang, Li and Yu, Bei and Sun, Qi and Zhuo, Cheng},
  title = {An agile framework for efficient {LLM} accelerator development and model inference},
  year = {2025},
  booktitle = {Proc. 43rd IEEE/ACM Int. Conf. Comput.-Aided Design},
  doi = {10.1145/3676536.3676753},
  no = {200},
  pages = {1--9}
}

@article{llm4fpga,
author = {Chen, Hongzheng and Zhang, Jiahao and Du, Yixiao and Xiang, Shaojie and Yue, Zichao and Zhang, Niansong and Cai, Yaohui and Zhang, Zhiru},
title = {Understanding the potential of {FPGA}-based spatial acceleration for large language model inference},
year = {2024},
volume = {18},
number = {1},
doi = {10.1145/3656177},
journal = {ACM Trans. Reconfigur. Technol. Syst.},
month = dec
}

@inproceedings{llama2fpga,
author = {Xu, Han and Wang, Xingyuan and Ji, Shihao},
title = {Towards energy-efficient {Llama2} architecture on embedded {FPGAs}},
year = {2024},
booktitle = {Proc. 33rd ACM Int. Conf. Inf. Knowl. Manage.},
pages = {5570-5571},
doi = {10.1145/3627673.3679068}
}

@INPROCEEDINGS{mecla,
  author={Qin, Yubin and Wang, Yang and Zhao, Zhiren and Yang, Xiaolong and Zhou, Yang and Wei, Shaojun and Hu, Yang and Yin, Shouyi},
  booktitle={Proc. 51st Annu. ACM/IEEE Int. Symp. Comput. Archit. (ISCA)},
  title="{MECLA}: Memory-compute-efficient {LLM} accelerator with scaling sub-matrix partition",
  year={2024},
  pages={1032-1047},
  doi={10.1109/ISCA59077.2024.00079}}

@INPROCEEDINGS{coro,
  author={Tao, Yongjin and Sun, Wendi and Chen, Song and Kang, Yi},
  booktitle={Proc. 17th IEEE Int. Conf. Solid-State Integr. Circuit Technol. (ICSICT)},
  title={Accelerating matrix-vector multiplications of large language models via efficient encoding},
  year={2024},
  pages={1-3},
  doi={10.1109/ICSICT62049.2024.10831638}}

@inproceedings{picachu,
author = {Qin, Jiajun and Xia, Tianhua and Tan, Cheng and Zhang, Jeff and Zhang, Sai Qian},
title = "{PICACHU}: Plug-in {CGRA} handling upcoming nonlinear operations in {LLMs}",
year = {2025},
booktitle = {Proc. 30th ACM Int. Conf. Archit. Support Program. Lang. Oper. Syst. (ASPLOS)},
pages = {845-861},
doi = {10.1145/3676641.3716013}
}

@misc{bitnet,
      title={{BitNet}: Scaling 1-bit Transformers for Large Language Models}, 
      author={Hongyu Wang and Shuming Ma and Li Dong and Shaohan Huang and Huaijie Wang and Lingxiao Ma and Fan Yang and Ruiping Wang and Yi Wu and Furu Wei},
      year={2023},
      eprint={2310.11453},
      archivePrefix={arXiv},
      primaryClass={cs.CL},
      note={arXiv:2310.11453}
}

@techreport{nvidia_h100,
  author = {{NVIDIA}},
  title = "{NVIDIA} {H100} {T}ensor {C}ore {GPU} architecture: Exceptional performance, scalability, and security for the data center",
  year = {2022},
  institution = {{NVIDIA}},
  address     = {Santa Clara, CA, USA},
}

@misc{sze2017efficientprocessingdeepneural,
      title={Efficient Processing of Deep Neural Networks: A Tutorial and Survey}, 
      author={Vivienne Sze and Yu-Hsin Chen and Tien-Ju Yang and Joel Emer},
      year={2017},
      note={arXiv:1703.09039},
      archivePrefix={arXiv},
      primaryClass={cs.CL}
}

@article{power_consideration,  
title={Energy and policy considerations for modern deep learning research},
volume={34},
doi={10.1609/aaai.v34i09.7123},
number={09},
journal={Proc. AAAI Conf. Artif. Intell.},
author={E. Strubell and A. Ganesh and A. McCallum},
year={2020},
month={Apr.},
pages={13693-13696}
}
